\documentclass[aps, prd, preprint, preprintnumbers, showpacs, showkeys, amsfonts, nofootinbib, 
floatfix,superscriptaddress] {revtex4-1}
\usepackage{graphicx}
\usepackage{epsfig}
\usepackage{rotate}
\usepackage{rotating}
\usepackage{multirow}
\usepackage[T1]{fontenc}
\usepackage{array}
\usepackage{graphicx,
}
\usepackage{epstopdf}
\usepackage{amsmath,amssymb,hyperref,enumerate}
\usepackage{pstricks}
\usepackage{slashed}
\usepackage{url}
\usepackage{color}
\hypersetup{colorlinks,citecolor= nicegreen,linkcolor= nicered}
\definecolor{nicered}{rgb}{0.7,0.1,0.1}
\definecolor{nicegreen}{rgb}{0.1,0.5,0.1}

\usepackage[normalem]{ulem}

\def\({\left(}
\def\){\right)}
\def\[{\left[}
\def\]{\right]}

\begin{document}
\preprint {RECAPP-HRI-2014-001; \, OSU-HEP-14-01}

\title{Searching for an elusive charged Higgs at the Large Hadron Collider}
\author{Ushoshi~Maitra}
\email{Email: ushoshi@hri.res.in}
\affiliation{Regional Centre for Accelerator-based Particle Physics, Harish-Chandra Research Institute,  Chhatnag Road, Jhusi, Allahabad 211019, India} 
\author{Biswarup~Mukhopadhyaya}
\email{Email: biswarup@hri.res.in}
\affiliation{Regional Centre for Accelerator-based Particle Physics, Harish-Chandra Research Institute,  Chhatnag Road, Jhusi, Allahabad 211019, India}
\author{S.~Nandi}
\email{Email: s.nandi@okstate.edu} 
\affiliation{Department of Physics and Oklahoma Center for High Energy Physics, 
Oklahoma State University, Stillwater OK 74078-3072, USA}
\author{Santosh~Kumar~Rai}
\email{Email: skrai@hri.res.in}
\affiliation{Regional Centre for Accelerator-based Particle Physics, Harish-Chandra Research Institute,  Chhatnag Road, Jhusi, Allahabad 211019, India}
\author{Ambresh~Shivaji}
\email{Email: ambreshkshivaji@hri.res.in}
\affiliation{Regional Centre for Accelerator-based Particle Physics, Harish-Chandra Research Institute,  Chhatnag Road, Jhusi, Allahabad 211019, India}

\date{\today}

\begin{abstract}
\noindent
We study the signals for a "fermiophobic" charged Higgs boson present in an extension of the 
standard model with an additional Higgs doublet and right handed neutrinos, responsible for 
generating Dirac-type neutrino masses. We study the pair production of the charged Higgs at the 
Large Hadron Collider (LHC), which can be relatively light and still allowed by experimental data. 
The charged Higgs decays dominantly into a $W$ boson and a very light neutral scalar present in 
the model, which decays invisibly and passes undetected. We find that the signal for such a charged 
Higgs is overwhelmed by the standard model background and will prove elusive at the 8 TeV run of the LHC. 
We present a cut-flow based analysis to pinpoint a search strategy at the 14 TeV run of the LHC which 
can achieve a signal significance of 5$\sigma$ for a given mass range of the charged Higgs.
\end{abstract}

\maketitle

\section{Introduction}
\label{sec:intro}
The search for new physics beyond the standard model is
continuing at the Large Hadron Collider (LHC). Although we are yet to 
see any clear hint, there is reason for
exhilaration in another way, namely, the discovery of a boson of mass
$\sim$ 125 GeV~\cite{atlascms}. The properties of this particle are very similar to
the Higgs boson predicted in the standard model (SM), but
possibilities of some new physics information contained in it cannot
yet be ruled out. Thus a great deal of attention has shifted to the
exploration of physics beyond the standard model (BSM) in the
electroweak symmetry breaking sector.

One vexing issue, often mentioned as a motivation for BSM physics, is
the identification of a mechanism for neutrino mass generation. This
basically means finding some explanation for the smallness of neutrino
masses as compared to those for the other fermions, and also the very
different nature of mixing evinced in the neutrino (or more precisely,
lepton) sector~\cite{GonzalezGarcia:2002dz}. It is thus natural that efforts to unravel new physics
in the Higgs sector will sometimes be guided by considerations related
to neutrino masses~\cite{Ma:2000cc,Gabriel:2006ns,Wang:2006jy,Davidson:2009ha}. The LHC signals of any scenario proposed in this
context are also of undeniable interest ~\cite{Davidson:2010sf,Gabriel:2008es}. 

In this paper, we consider a model which not only accounts for the tiny neutrino masses, 
but also plays a role in the electroweak symmetry breaking mechanism. We are interested 
in a two Higgs doublet model with right-handed neutrinos proposed by Gabriel and 
Nandi~\cite{Gabriel:2006ns}.  The essential idea is that neutrinos, like all other fermions, have 
Dirac masses, but are  much lighter than the others because their masses come from a 
different Higgs doublet. The Yukawa couplings of the neutrinos can still be
${\cal O}(1)$.  This is ensured by giving a very tiny vacuum expectation value (vev) $\sim$ eV 
to the neutral component of one of the Higgs doublets, which due to a $Z_2$ symmetry couples 
only with the neutrino sector. The charged Higgs in this model therefore has very different properties 
when compared to other standard two Higgs doublet models (2HDM) \cite{Branco:2011iw}. It is found to couple very 
weakly with the quarks while a large coupling with charged leptons and right-handed neutrinos is allowed. 
Thus, the leptonic mode is the most promising decay channel of this charged Higgs for ${\cal O}(1)$ 
Yukawa couplings in the neutrino sector  \cite{Gabriel:2008es}.  Depending upon the mass of the 
charged Higgs other decay modes are also possible. In a similar model proposed by Davidson and 
Logan \cite{Davidson:2009ha,Davidson:2010sf}, which considers a $U(1)$ global symmetry in place of the discrete $Z_2$ symmetry, 
only leptonic decay modes of the charged Higgs were allowed.

The branching probabilities for the decay of the charged Higgs are very sensitive to 
the small vev of the additional doublet, which couples to the
neutrinos \cite{Gabriel:2008es}. As recent astrophysical bounds on the neutrino Yukawa 
couplings in such extensions of the SM suggest that the vev of the second doublet cannot be 
in the sub-eV range \cite{Davidson:2009ha,Zhou:2011rc}, the charged Higgs can no longer decay 
dominantly into a leptonic final state. It turns out that the main decay mode for the charged Higgs 
is into a $W$ boson and a light neutral scalar present in the model. In this work we mainly focus on the 
challenges presented for a charged Higgs search at the LHC, when it decays through the $W$ mode. We 
study the pair production of the charged Higgs at the LHC and consider its decay to $W$ boson and the 
neutral scalar. The signal is identified by two isolated leptons and a large missing energy. We have 
analyzed the most dominant SM background subprocesses that affect the signal, to estimate the signal significance. 
The study is carried out at both the 8 TeV and 14 TeV center-of-mass energies for the LHC.  We
find that it is practically impossible to achieve any significant signal for the charged Higgs in this model
with the available integrated luminosity at 8 TeV. The situation is much more optimistic at 14 TeV, if one waits 
for a sizeable luminosity to accumulate. There, on applying appropriate kinematical event selection
procedure, a signal significance of 5$\sigma$ at 14 TeV can be reached with an integrated luminosity of 
4000 fb$^{-1}$ for $M_{H^\pm} = 150$ GeV. The observation of the signal is found to be more promising for heavier 
charged Higgs masses.

The paper is organized as follows. In Section \ref{sec:model} we describe the model briefly. 
Various constraints on the model parameters are discussed in Section \ref{sec:constraints}. 
In Section \ref{sec:analysis}, we discuss the charged Higgs pair production and major backgrounds for
the signal at the LHC. Results are presented in Section \ref{sec:results} and we
conclude in Section \ref{sec:summary}.
 
 \section{A brief review of the Model} 
 \label{sec:model}
 The model under consideration is based on the symmetry group $\mathcal{G}_{SM}\times Z_2$, where
 $\mathcal{G}_{SM}\equiv SU(3)_c\times SU(2)_L\times U(1)_Y$. In addition to the matter fields in the SM, the model 
 includes two scalar doublets $\chi$ and $\phi$, and three $SU(2)_L$ singlet right-handed neutrinos 
 $\nu_R^i$, $i=1,2,3$. All the SM fermions and the scalar doublet $\chi$ are even under the
 discrete symmetry $Z_2$, while the right-handed neutrinos and the
 scalar doublet $\phi$ are odd under $Z_2$. The most general scalar potential and the 
Yukawa interaction of leptons with the scalar doublets which respect the $\mathcal{G}_{SM}\times Z_2$ symmetry are \cite{Gabriel:2006ns},
   \begin{eqnarray}
{\cal V} &=& -\mu_1^2\;\chi^\dagger \chi -\mu_2^2\;\phi^\dagger \phi + \lambda_1\;(\chi^\dagger \chi)^2
   + \lambda_2\;(\phi^\dagger \phi)^2  + \lambda_3\;(\chi^\dagger \chi)(\phi^\dagger \phi) \nonumber \\
   && -\lambda_4|(\chi^\dagger \phi)|^2 
      -\frac{1}{2} \lambda_5 \left[ (\chi^\dagger \phi)^2 + (\phi^\dagger \chi)^2 \right], \label{eq:pot} \\
{\cal L}_Y &=&  y_l^{ij}\; {\bar{\rm \Psi}}^{l,i}_L l_R^j \chi + y_{\nu_l}^{ij}\; {\bar{\rm \Psi}}^{l,i}_L \nu_R^j {\tilde \phi} + h.c.,
\label{eq:Lyuk}
   \end{eqnarray}
   where, ${\bar{\rm \Psi}}^{l,i}_L = ({\bar \nu}_l^i,{\bar l}^i)_L$
   and $l_R^j$ are the usual $SU(2)_L$ lepton doublet and singlet
   fields, respectively and $y_f^{ij}~(f \equiv l, \nu_l)$ represent the matrix elements of the 
   lepton Yukawa matrices. The standard electroweak symmetry is
   broken spontaneously by giving a vev, $V_\chi \simeq$ 246 GeV to
   the $\chi$ doublet, while the $Z_2$ symmetry is broken by a vev,
   $V_\phi$ for the $\phi$ doublet. The spontaneous breaking of the
   $Z_2$ symmetry is arranged for generating small neutrino masses,
   $m_{\nu_l} \sim$ $V_\phi$ which can be in the sub-eV/eV range for ${\cal O}(1)$ 
   Yukawa couplings. We note that we are assuming lepton number conservation so that
the Majorana mass terms  for the right handed neutrinos, $\nu_R$,
$M  \nu_R ^T C^{-1} \nu_R$  are not allowed. Thus the light left-handed
neutrinos cannot acquire masses via the usual see-saw mechanism ~\cite{Mohapatra:1979ia}.
Dirac mass, as obtained from Eq.~\ref{eq:Lyuk}  from the tiny vev of $\phi$ is the
only possibility.
   
   As a result of the symmetry breaking, the physical Higgs sector
   includes charged scalars $H^\pm$, two neutral CP-even scalars $h$ and
   $\sigma$ and a neutral pseudoscalar $\rho$.  The masses for these particles are given by,
   \begin{align}
    & M_{H^\pm}^2 = \frac{1}{2} (\lambda_4 + \lambda_5) V^2, &&  M_\rho^2 = \lambda_5 V^2 \nonumber \\
    & M_h^2 = 2 \lambda_1 V_\chi^2, &&   M_\sigma^2 = 2 \lambda_2 V_\phi^2, 
    \label{eq:masses}
   \end{align}
   where, $V^2 = V_\chi^2 + V_\phi^2$. We have neglected the
   subdominant terms in $V_\phi$ when deriving these relations. 
We note that in the case of exact $Z_2$ symmetry, the  $\sigma$ will
be exactly massless. The  breaking of this $Z_2$ symmetry with a tiny vev
$ V_{\phi}$ gives mass to the $\sigma$, as well as tiny Dirac masses to
the observed neutrinos. 
Therefore in this model, the neutral scalar field $\sigma$ 
   is very light and the field $h$ behaves like the SM Higgs boson.The CP-even scalars 
   ($h,\sigma$) are the mass eigenstates and they are related to the weak eigenstates $(h_0,\sigma_0)$ 
    by the mixing angle $\theta$: 
   \begin{align}
   & h = h_0 \cos\theta -  \sigma_0 \sin\theta, &&\sigma = h_0 \sin\theta  + \sigma_0 \cos\theta .  
   \end{align}
   where,    
   \begin{align}
    & \cos\theta = 1 + {\cal O}(V_\phi^2/V_\chi^2) ,
    && \sin\theta = -\frac{\lambda_3-\lambda_4-\lambda_5}{2\lambda_1} \left( \frac{V_\phi}{V_\chi}\right) + {\cal O}(V_\phi^2/V_\chi^2). 
   \end{align}
This mixing can be neglected because $V_\phi << V_\chi$.
It is also clear from the above  equations that $M_\rho$ lies around the 
electroweak scale.

   In the lepton Yukawa sector the above symmetry breaking leads to
   neutrino masses given by, $m_{\nu_a} = y_{\nu}^a V_\phi/\sqrt{2}$,
   where $y_{\nu}^a$ are the eigenvalues of the neutrino Yukawa
   matrix. The Yukawa interaction of the charged Higgs with the leptons can then be written down following 
   Eq.~\ref{eq:Lyuk} as,
   \begin{equation}
    {\cal L}_Y \supset y_{\nu}^a \frac{V_\chi}{V}\; U_{ia} {\bar{l}}^i_L \nu_R^aH^- + y_l^i \frac{V_\phi}{V}\; U_{ia} {\bar{l}_R}^i \nu_L^a H^- + h.c.
    \label{eq:Yukawa}
   \end{equation}
   In the above equation, $i$ represents the flavour index while $a$ is the index
   representing neutrino components in mass eigenstate. The $y_l^i =
   \sqrt{2} m_l^i/V_\chi$ are the charged lepton Yukawa couplings while
   $U_{ia}$ represent the elements in the PMNS matrix~\cite{PMNS} for the mixing of the 
   neutrino flavours.\footnote{$\nu^i = \sum_a U_{ia} \nu^a $.}  Note that the 
   second term in Eq.~\ref{eq:Yukawa} is clearly sub-dominant and negligible 
  (suppressed by the factor $V_\phi/V$) when compared to the first term and can 
  therefore be safely neglected when considering the interaction strength of the charged 
  Higgs with the leptons. As the couplings of charged Higgs with SM
   quarks are also generated through terms similar to the second term in 
   Eq.~\ref{eq:Yukawa}, the charged Higgs is very weakly coupled to quarks and
   in all practicality behaves as a "chromophobic" field. This property of the 
 charged Higgs plays a crucial role in avoiding strong constraints on its mass, otherwise 
 evident in other 2HDM, from low energy physics experiments such as weak meson decays 
 and mixing. Thus in the Yukawa sector, only the decay $H^\pm \to l_L^\pm \nu_R$ becomes 
  relevant. Other main decay modes of the charged Higgs include  $H^\pm \to W^\pm \sigma$ 
  and $H^\pm \to W^\pm \rho$ which have gauge coupling interaction strengths. 
\begin{figure}[h!]
\includegraphics[width=3.15in]{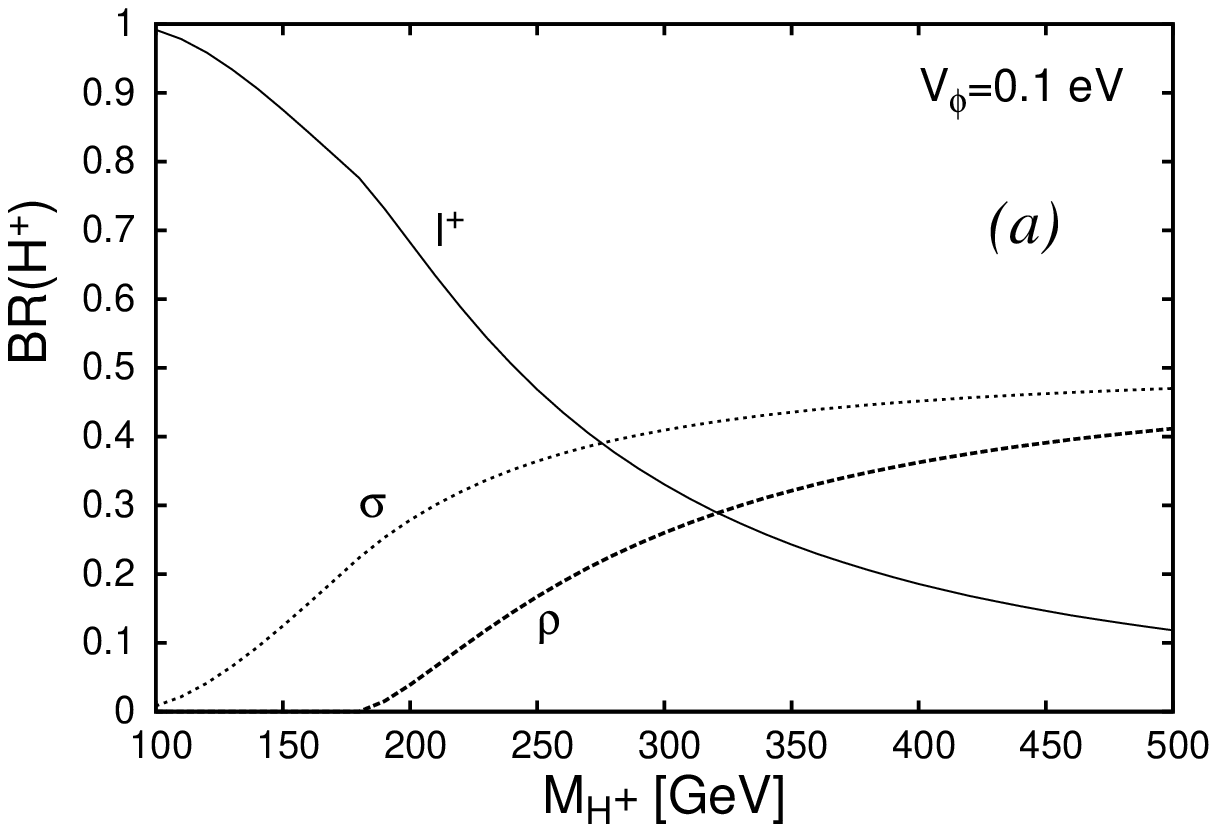}
\includegraphics[width=3.15in]{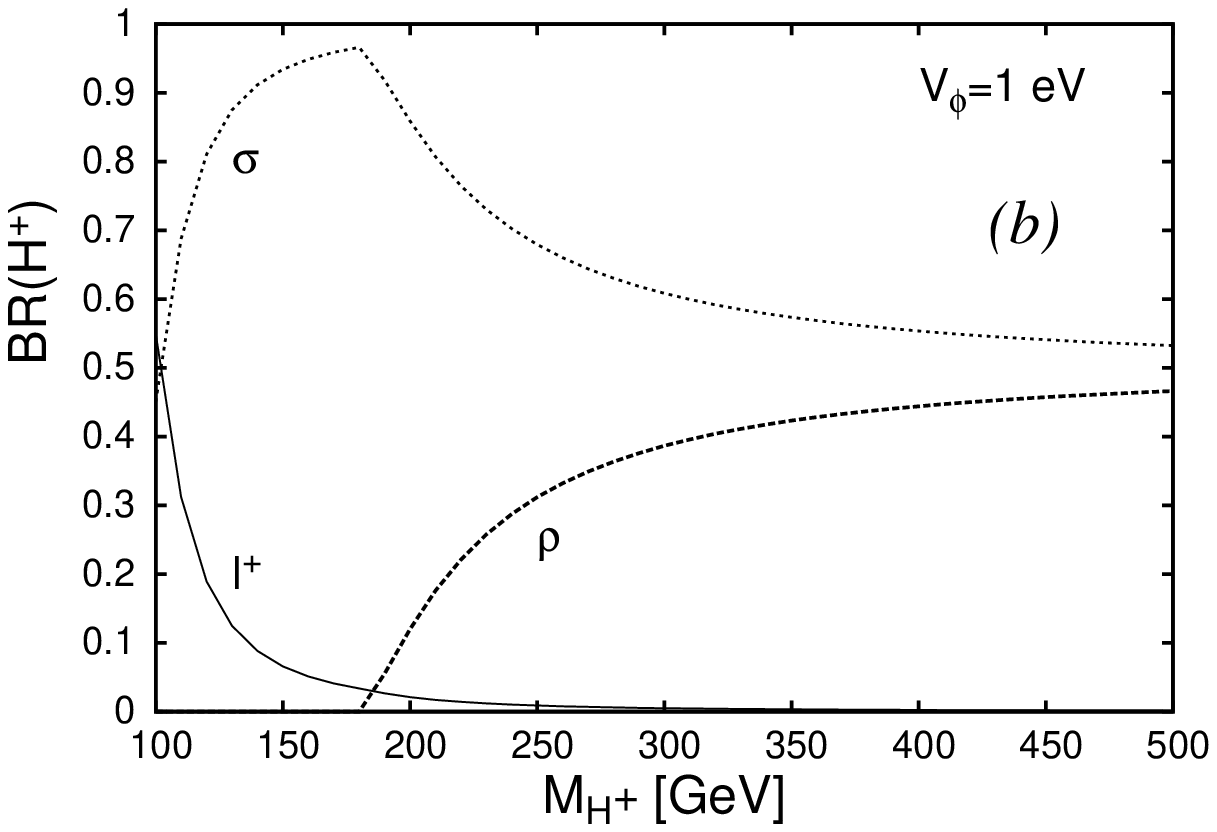}
\caption{\small Charged Higgs branching ratios as function of its mass for (a) $V_\phi = 0.1$ eV  and 
(b) $V_\phi = 1$ eV. The mass of $\sigma$ particle is related to $V_\phi$ as in Eq.~\ref{eq:masses}. 
We have chosen the other relevant variables $\lambda_2 = 1.0$ and $M_\rho = 100$ GeV for calculating 
the above branching ratios.}
\label{fig:BR}
\end{figure}
  The branching ratios of charged Higgs decay is quite sensitive to the value of $V_\phi$. As the 
   leptonic channel $(l~\nu_l)$ is dictated by the coupling strength given by 
   $\sim m_\nu/V_\phi$ (see Eq.~\ref{eq:Yukawa}), smaller values of $V_\phi$ for a fixed neutrino mass
  increase the branching probability. To highlight this, we consider two sets of values for $V_\phi$ 
  and show the branching probabilities of the charged Higgs decay as a function of its mass in 
   Fig.~\ref{fig:BR}. Note that the neutrino data as shown in Table~\ref{Table:nudata} have been incorporated
   when calculating the partial decay widths of the charged Higgs decaying into the three generations of 
leptons. If $V_\phi$ is in the sub-eV range as shown in Fig.~\ref{fig:BR}a, it is found to decay mostly 
through the leptonic mode for $M_{H^\pm} \leq 200$ GeV, while if $V_\phi$ is increased to about an eV 
(Fig.~\ref{fig:BR}b), it decays dominantly into $W^{\pm}\sigma$. As the value of $V_\phi$ is increased further, 
we find that the leptonic channel becomes completely negligible and the $W^{\pm}\sigma$ becomes the only 
significant mode available for the charged Higgs decay for $M_{H^\pm} \leq 200$ GeV.

 \section{Constraints on Model Parameters }
 \label{sec:constraints}
As the model under consideration is quite different from the generic 2HDM and is envisioned to account for
the observed neutrino masses and mixing angles, it becomes imperative to first check how the experimental 
constraints affect the parameters of the model. A brief discussion on several constraints on the model 
parameters is already present in Ref. \cite{Gabriel:2006ns}. We choose to accommodate them with new 
and updated results that have modified these constraints as well as supplement them with additional 
constraints, if any. 

 \begin{table}[h!]
 \begin{center}
 \begin{tabular}{|c c c|}
   \hline
  Parameters & NH & IH   \\
 \hline
 ${\rm sin}^2\theta_{12}$ & $0.307^{+0.052}_{-0.048}$ & $0.307^{+0.052}_{-0.048}$ \\ 
 ${\rm sin}^2\theta_{23}$ &  $0.386^{+0.055}_{-0.251}$ &  $0.392^{+0.057}_{-0.271}$ \\
 ${\rm sin}^2\theta_{13}$ &  $2.41^{+0.72}_{-0.72} \times 10^{-2}$ &  $2.42^{+0.73}_{-0.71} \times 10^{-2}$ \\
 $m_{\nu_2}^2-m_{\nu_1}^2$ & $7.54^{+0.55}_{-0.64} \times 10^{-5} {\rm eV}^2$ & $7.54^{+0.55}_{-0.64} \times 10^{-5} {\rm eV}^2$\\
 $m_{\nu_3}^2-\frac{1}{2}( m_{\nu_1}^2 + m_{\nu_2}^2)$ & $2.43^{+0.24}_{-0.19} \times 10^{-3} {\rm eV}^2$ & $-2.42^{+0.25}_{-0.19} \times 10^{-3} {\rm eV}^2$\\
  \hline
 \end{tabular}
 \end{center}
 \caption{Neutrino mass-mixing parameters with 3$\sigma$ uncertainties~\cite{Fogli:2012ua}. The allowed ranges 
          of parameters for the Normal Hierarchy (NH) and Inverted Hierarchy (IH) cases are shown separately.}
\label{Table:nudata}
\end{table}
We acknowledge that any scenario explaining neutrino masses will also need to address their mixing, and reproduce the 
form of the PMNS matrix as suggested by various observations~\cite{Fogli:2012ua}.
The PMNS matrix is parameterized by three mixing angles and can have one phase whose value is yet unknown.
 The current values of these angles and the neutrino
 mass-squared differences are shown in  Table~\ref{Table:nudata}.
From the measurements on the neutrino mass-squared differences we can
conclude that in both the normal and inverted hierarchy scenarios, the
mass of the heaviest neutrino is $\gtrsim$ 0.05 eV. We have already
discussed the sensitivity of the branching ratios of the charged Higgs to the magnitude of $V_\phi$. 
Clearly, from neutrino data, one is free to choose ${\cal O}(1)$ Yukawa couplings ($y_\nu$).   
However, the right-handed neutrinos are new relativistic degrees of freedom present 
in our model. Due to its coupling with the charged Higgs and the leptons, 
they should be excessively produced in the early Universe, for example, via the 
charged Higgs mediated $t-$channel process $l^+l^- \leftrightarrow \nu_R {\bar \nu}_R$. 
We can therefore put a constraint on the neutrino Yukawa coupling using the big-bang 
nucleosynthesis (BBN) bound on new relativistic degrees of freedom ($\delta N_\nu$) \cite{Steigman:1979xp}.
The latest results combining {\tt Planck}, {\tt WP}, Baryon Acoustic Oscillation ({\tt BAO}) and high multipole {\tt CMB} data, on the upper limit of extra relativistic degrees of freedom give
$\delta N_{\nu,max} \simeq 1.0$ at 95\% confidence level \cite{Ade:2013ktc} .
 \begin{figure}[t!]
\includegraphics[width=3.15in]{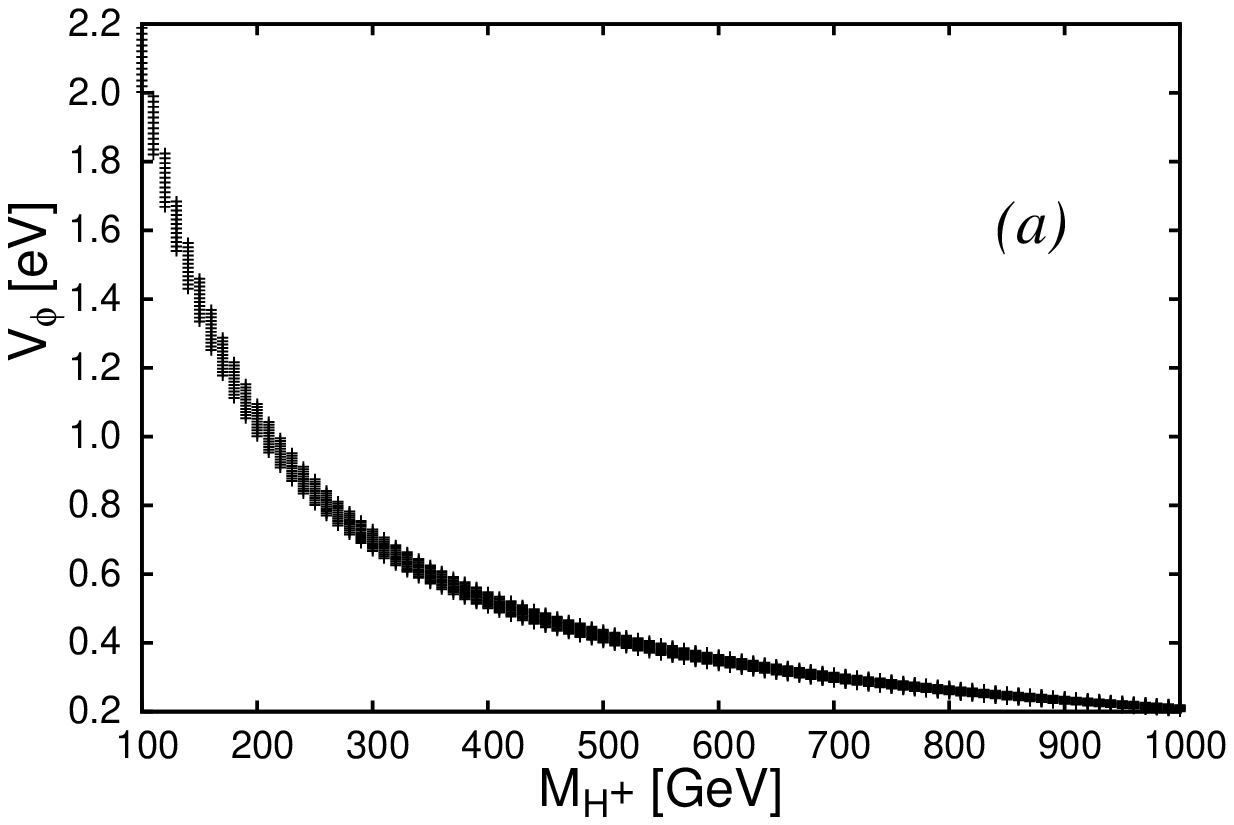}
\includegraphics[width=3.15in]{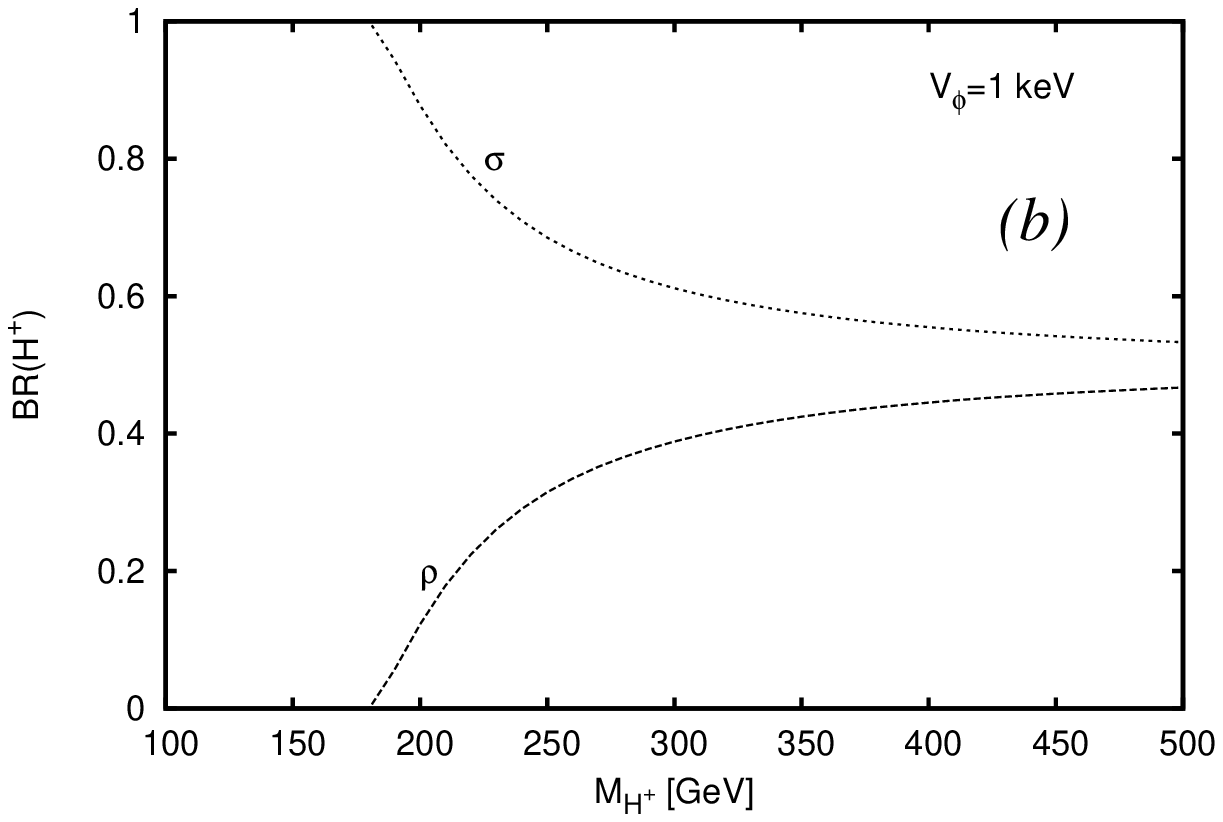}
\caption{(a)  The variation of the lower bound on $V_{\phi}$ as a function of $M_{H^{+}}$ as defined by 
Eq.~\ref{eq:bbnbound}. The band represents the $3\sigma$ 
uncertainties shown in Table~\ref{Table:nudata}. (b) The branching ratios of the charged Higgs decay for 
the choice of $V_\phi=1$ keV.}
\label{fig:scanplot}
\end{figure}%
This bound can be translated into an upper bound on the neutrino Yukawa coupling or a 
lower bound on $V_\phi$ for a given neutrino mass which is given by ~\cite{Davidson:2009ha}
  \begin{equation}
  V_{\phi} \gtrsim 60\;m_{\nu_{i}}(|U_{li}|)\frac{100\; {\rm GeV}}{M_{H^{+}}}.
 \label{eq:bbnbound}
  \end{equation}
The lower bound on $V_\phi$ is derived for the most massive neutrino
labeled by $l$ in $U_{li}$. The hierarchy in the neutrino masses is
therefore not important here.  We have considered the maximal mixing
for which $ |U_{li}| \simeq 1/\sqrt{2}$. A value of $ U_{li} $
consistent with the neutrino data does not alter the numerical value
of the bound significantly. If $M_{H^\pm} \sim 100$ GeV, the above bound
implies $V_\phi \gtrsim 2$ eV. In Fig.~\ref{fig:scanplot}a, we
have shown the variation of the lower bound on $V_{\phi}$ as a function of 
$M_{H^\pm}$ consistent with the neutrino data. 
For a fixed value of $M_{H^\pm}$, the range of $V_\phi$ illustrates $3\sigma$
uncertainty in neutrino data.
This, when considered with 
the decay properties of the charged Higgs illustrated in Fig.~\ref{fig:BR} shows that 
for a light charged Higgs (100-200 GeV), the dominant decay is to $W\sigma$ as $V_\phi > 1$ eV. 

In addition to that, if the supernova neutrino
observations and the energy-loss argument for supernova cores are also
considered, the lower bound on $V_\phi$ can be pushed to $V_\phi
\gtrsim 1$ keV~\cite{Zhou:2011rc}. This also takes care of the excessive 
production of neutrinos through the $\sigma$ boson mediated process
$\nu_i + {\bar \nu}_i \to \nu_j + {\bar \nu}_j$, in the early Universe.
Since $\Gamma(H^\pm \to l_L^\pm \nu_R) \propto m_{\nu}^2/V_\phi^2$, the
leptonic decay mode of the charged Higgs will be extremely suppressed as the Yukawa couplings 
are further suppressed ($y_\nu \sim m_\nu/V_\phi$). Instead, it will decay overwhelmingly via the 
modes $H^{\pm}\to W^{\pm}\sigma$ and $W^{\pm}\rho$ (see Fig.~\ref{fig:scanplot}b) , thus behaving 
more like a "fermiophobic" field. 
 
The values of coupling parameters $\lambda_1,\lambda_2,\lambda_4$ and
$\lambda_5$ appearing in the scalar potential (Eq.~\ref{eq:pot}) can be fixed once we 
make a choice for the scalar masses $M_h, M_\sigma, M_\rho$ and $M_{H^\pm}$ (see Eq.~\ref{eq:masses}). 
To incorporate the recently discovered SM-like Higgs boson in our model, we would like to have
$M_h \sim 125-126$ GeV, which fixes $\lambda_1$.  We can choose any value for $\lambda_2$ which
is not very large so that $M_\sigma \simeq V_\phi \sim $ keV.  Note that a 1 keV
$\sigma$ particle will decay into neutrinos in about $10^{-9}$ seconds.  Although 
${\cal O}(1)$ values of $\lambda_3$ do not affect the scalar mass spectrum, it appears in various 
interaction vertices.
We find that by choosing $\lambda_3 = \lambda_4+\lambda_5$ we can suppress large contributions to
the invisible decay width of the SM-like Higgs via the decay mode $ h \to \sigma \sigma$.  
As the present LHC data allows a maximum of $20 \%$ branching ratio for any
invisible decay mode(s) of the Higgs boson at the 95 $\%$ C.L.~\cite{ATLAS:2013,CMS:2013}, allowing
$BR(h \to \sigma \sigma) = 20\%$ puts a condition on $\lambda_3$ which is
given by,
  \begin{eqnarray}
      \lambda_3 &=& 0.0133 + \lambda_5 + \lambda_4, \nonumber \\
                &=& 0.0133 + 0.3305 \times \left(\frac{M_{H^\pm}}{100\; \rm GeV}\right)^2,
\label{eq:l3}
  \end{eqnarray}
where we have used Eq.~\ref{eq:masses} and $V \simeq V_\chi =$ 246 GeV.
Even with the above choice of $\lambda_3$, a light enough $\rho$ may further
  contribute to the Higgs invisible decay width.  We may therefore
  choose $M_\rho$ sufficiently large so that this situation is avoided.
   The pseudoscalar $\rho$, belonging to the doublet $\phi$ has no significant interaction
  with charged leptons and quarks, and decays mostly into a neutrino-antineutrino
  pair. Since the decay $Z \to \rho \sigma$ contributes to the invisible decay width
  of the $Z$ boson, the experimental measurements require  $M_\rho
  \gtrsim$ 78 GeV \cite{pdg}.  When $M_\rho > m_Z$ one also needs to consider the LEP2 data for 
  the signal from the process $e^+e^- \to Z^\ast \to \rho \sigma$. Non observation of any such
 signal puts a lower bound on the $\rho$ mass of 95 GeV \cite{Gabriel:2006ns}. Note that if we take 
 $M_\rho = 100\; {\rm GeV}$, $\Gamma(h  \to \rho \rho^*) \sim {\rm eV}$  which will have negligible contribution to the Higgs invisible decay width.

 The only bound on the mass of the charged Higgs in our model comes
  from the direct searches for the pair production at the LEP
  experiments.  Due to  very suppressed coupling of the charged Higgs with the quarks, the
  constraints from rare processes such as $b \to s \gamma$ do not put
  any additional bound on the charged Higgs mass. Thus it is enough to have
  $m_{H^\pm} \ge$ 79.3 GeV \cite{pdg}.
   
\section{Prospects of the charged Higgs at the LHC: Signal background analysis}
 \label{sec:analysis}

 \begin{figure}[t]
  \includegraphics[width=4.2in]{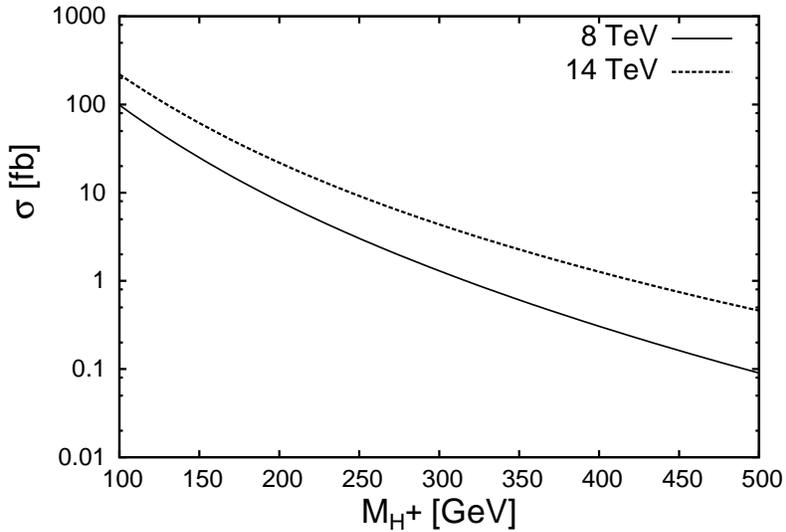}
   \caption{ Charged Higgs pair production cross section as function of charged Higgs mass at 8 TeV 
   and 14 TeV LHC center-of-mass energies.}
   \label{fig:xsplot}
\end{figure}%
Since the chromophobic nature of the charged Higgs in this scenario
disallows its production in association with a top (anti-top) quark, one has to rely on electroweak 
sub-processes for its pair production. Thus the $H^{\pm}$ pair is produced via Drell-Yan
process through the exchange of photon and $Z$ boson in the s-channel. It can also be produced at the 
LHC through vector boson fusion (VBF), namely, $ qq \to qqV^*V^* \to qqH^{+}H^{-}$ where 
$V = \gamma,Z,W^{\pm}$. However, the VBF production cross section is quite suppressed when 
compared to Drell-Yan. The pair production cross section for the charged Higgs at the LHC at 
center-of-mass energies of 8 TeV and 14 TeV is shown in Fig.~\ref{fig:xsplot}. The cross sections have been computed using {\tt CTEQ6L1} parton distribution functions (PDFs) \cite{Pumplin:2002vw}.
Since the coupling of 
SM-like Higgs boson with the charged Higgs is not negligible, the charged Higgs pair production may 
also receive additional contributions via a Higgs ($h$)  mediated gluon fusion process. For our choice 
of $\lambda_3$ and $M_{H^\pm} = 150$ GeV, the $gg \to  h^* \to H^+ H^-$, we 
find the gluon mediated cross section is about 1.86 fb for the 14 TeV run of the LHC.  For larger values  
of $M_{H^\pm}$, one expects this contribution to grow, as $\lambda_3$ also 
increases (Eq.~\ref{eq:l3}). But the s-channel mediated process receives a significant propagator 
suppression (as the effective $\hat{s} >  2M_{H^{\pm}}$ for pair production), making it quite small for 
larger $M_{H^\pm}$.

The charged Higgs can also be produced singly in association with a
$\rho$ or $\sigma$ in the channel $ qq' \to \rho H^{\pm} , \sigma
H^{\pm} $. These production modes lead to single $H^{\pm}$ along with
large missing energy when $\rho$ and $\sigma$ decay to $\nu
\bar{\nu}$. Although the rate of single charged Higgs production is
comparable to that of pair produced charged Higgs, the single-$W^\pm$ 
background is very large compared to that from $W^+W^-$. So we study the signals
for the charged Higgs via pair production.
As illustrated in Fig.~\ref{fig:scanplot}b when $V_{\phi} \sim {\rm
  keV}$, $H^{\pm} \to W^{\pm} \sigma$ is the most favourable decay
channel as compared to other decays.
As $\sigma$ decays to neutrinos with 100\% branching probability, we will be focusing on
events with large transverse missing energy ($\slashed{E}_T$). The $W$ produced
from the charged Higgs, decays into $(l~\nu_l)$. Thus, the signature of
charged Higgs in this model is $pp \to H^{+}H^{-} \to W^{+}W^{-} \sigma \sigma \to l^{+}l^{-} + \slashed{E}_T$.

For our analysis, we have used the package {\tt MadGraph 5} \cite{Alwall:2011uj} to generate events for the 
signal as well as the SM background processes. To generate the events for the signal, we have included the 
interaction vertices of the new model in {\tt MadGraph 5}  using the publicly available package 
{\tt Feynrules} \cite{Christensen:2008py}. We have kept the factorization and renormalization scales same as
the default event-by-event {\tt MadGraph 5} value which happens to be the transverse mass of the pair produced particle ~\cite{scale:mg}.
A full simulation of the generated events has been carried out
by including fragmentation and hadronization effects using {\tt PYTHIA 8} \cite{Sjostrand:2007gs}.
We also include the  initial and final state radiations. In order to get a real assessment of the signal and 
the background at the detector level we have considered isolated leptons and jets. The event selection criteria 
that we use is consistent with that of the ATLAS detector \cite{ATLAS:2012mec}. However, a 100\% lepton 
identification (for $e$ and $\mu$) efficiency is assumed. To account for the detector resolutions we have smeared the energies/transverse momenta of leptons and jets with Gaussian functions as shown in Table \ref{Table:smearing} \cite{Aad:2009wy}. 
\begin{table}[ht!]
 \begin{center}
  \begin{tabular}{|c|c|c|c|c|}
   \hline
  {}& {\bf Electrons} & {\bf Muons}&  {\bf Jets} & {\bf Uncl. Energy}     \\

	      & $\frac{\sigma(E)}{E}$ & $\frac{\sigma(p_{T})}{p_{T}}$ & $\frac{\sigma(E)}{E}$ & $\sigma(E)$ \\
  \hline
  Formula      & $\frac{a}{\sqrt{E}} \oplus b \oplus \frac{c}{E}$ & $ a $ if $p_{T} < 100$ & $\frac{a}{\sqrt{E}} \oplus b \oplus \frac{c}{E}$& $\alpha \sqrt{\sum_{i}E_{T}^{uncl}}$\\
		&     &  else $a + b\log(\frac{p_{T}}{100})$ &   &   \\
\hline
  $|\eta| < 1.5$ & $a = 0.11, \; b = 0.007, $  & $a = 0.02, \; b = 0.08$  & $a = 0.65, \; b = 0.027,$& $\alpha = 0.55$\\
                 & $ c = 0.25 $& & $c = 4$  & \\
   \hline
 $ 1.5 < |\eta| < 2.5$ & $a = 0.13, \; b = 0.007,$  & $a = 0.03, \; b = 0.06$  & $a = 1.10, \; b = 0.01,$ & $\alpha = 0.55$ \\
		& $c = 0.25$ & & $c = 6.5$ & \\
\hline
  $2.5 < |\eta| < 3.0$ & -----  & -----  & $a = 1.10, \; b = 0.01,$ & $\alpha = 0.55$ \\
    & & & $c = 6.5$ & \\
   \hline	
  $3.0 < |\eta| < 4.5$ & -----  & -----  & $a = 1.00, \; b = 0.05,$ & $\alpha = 0.55$ \\
    &  &  & $ c = 1.0$ & \\
  \hline
  \end{tabular}
 \end{center}
 \caption{Functional form and parameters of the resolution functions of different physics objects. 
          These parameterizations give the value of $\sigma$ parameter of the gaussian functions used.
          The first and second column of the last two rows are kept blank as the leptons are identified 
          within $|\eta| < 2.5$. }
\label{Table:smearing}
\end{table}
  
The model parameters used in our analysis are
  \begin{align}\label{eq:parameters}
    \lambda_{1} &= 0.13,  & \lambda_{2} &= 1.0, & \lambda_{3} &= 0.0133 + \lambda_4 + \lambda_5, & \nonumber \\
   \lambda_{4} &= 2\frac{M_{H^{+}}^2}{V^2} - \lambda_{5}, & \lambda_{5} &= \frac{M_{\rho}^2}{V^2}, &  V_\phi &=1~{\rm keV},  &  \nonumber \\
   M_{\rho}   &= 100 ~{\rm GeV},  &  M_{\sigma} &= \sqrt2 V_{\phi}, &  
   M_{h}  &= 126~{\rm GeV}.  &
  \end{align}
It is worth pointing out that we have chosen neutrino masses (normal hierarchy) which are consistent with the neutrino data. 
In our analysis, however, neutrino masses and their hierarchy are of no consequence because 
the leptonic  decay mode of the charged Higgs for $V_\phi =1\; $keV is highly suppressed.
The major subprocesses in the SM that contribute as background to our signal are 
$pp \to t\bar{t}, ~W^{+}W^{-}, ~ZZ$ and also $pp \to h \to WW^*/ZZ^*$. Note that we have identified 
the Higgs ($h$) mediated subprocesses separately. As the Higgs production through gluon-fusion is a loop mediated process, 
we have included it  in {\tt Madgraph 5} via an effective operator. The $t\bar{t}$ 
background is a reducible background which can be ignored by selecting zero jet events. By removing 
the $Z$-peak (selecting a narrow window of 30 GeV around the peak in the invariant mass distribution
of the dileptonic system) we can also suppress the $ZZ$ background. This invariant mass cut partially 
takes care of the $pp \to h \to ZZ^*$ background when the on-shell $Z$ decays into charged lepton pair.  
As we shall explain in the next section, a large missing transverse energy ($\slashed{E}_T$) cut is essential 
for our signal-background analysis. We find that on applying a large $\slashed{E}_T$ cut the $pp \to h \to WW^*/ZZ^*$ 
backgrounds become negligible.  Thus the $ pp \to W^{+}W^{-}$ is the major irreducible background to our signal. 

Note that we have performed a leading order analysis here.  
Since the production of the charged Higgs takes place via a Drell-Yan process, QCD 
corrections are not expected to make any significant difference to the kinematic distributions, to be 
discussed later, on which our conclusions hinge so crucially.
 
\section{Results}
 \label{sec:results}
In this section, we present our results for charged Higgs masses of 150 GeV and 200 GeV as benchmark 
values. For $M_{H^\pm} = $ 150 (200) GeV, the pair production ($pp \to H^+H^-$) cross sections at 8 TeV 
and 14 TeV LHC center-of-mass energies are 21.48 (6.86) fb and 53.05 (18.73) fb respectively. The dominant  
decay mode of the charged Higgs for our choice of parameters is $W^\pm \sigma$ . The branching fraction of 
this decay mode for the charged Higgs mass of 150 GeV is close to 100\%. 
Since we have taken $M_\rho=100$ GeV, for 200 GeV charged Higgs the $W^\pm \rho$ decay mode is 
also allowed and $ H^\pm \to W^\pm \sigma$ branching probability reduces to about 88\% (see Fig.~\ref{fig:scanplot}b).  
Thus for the 200 GeV charged Higgs, $ W \sigma$ mode still remains the most dominant channel. However, even the $W^\pm \rho$ 
mode might contribute to our signal, since the $\rho$ can also decay invisibly. For $M_\rho=100$ GeV, there are two 
possible decay modes, {\it viz.} $\rho \to Z\sigma$ and $\rho \to \nu\bar{\nu}$. But much like the charged Higgs, the choice 
of $V_\phi=1$ keV suppresses the invisible decay of $\rho$ and it decays to $Z\sigma$ with 100\% branching probability. The 
decay of the charged Higgs is followed by the leptonic decay of $W$ boson. Since we have isolated both the leptons and jets, 
the events with isolated jets are removed and we select 
the signal and background events consisting of two isolated charged leptons and missing energy.

\begin{figure}[ht!]
\includegraphics[width=4.2in]{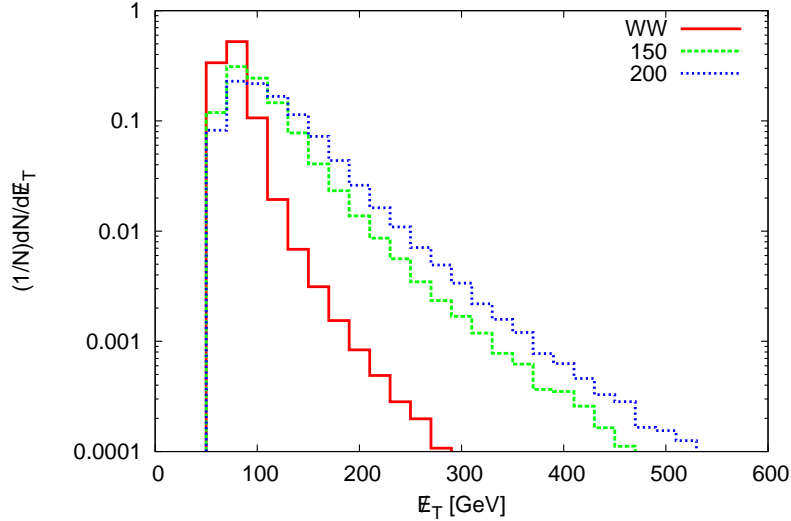}
 \caption{A comparison of the missing energy distributions for the signal and the $W^{+}W^{-}$ 
background at 8 TeV. Both the 150 GeV and 200 GeV charged Higgs mass cases of the signal is considered.}
  \label{fig:met}
\end{figure}

The basic acceptance cuts for the signal as well as background include,
\begin{equation}
 {\color{black} p_T^l > 20\; {\rm GeV}, \; |\eta_l| < 2, \; \Delta R_{ll} > 0.4, \; 
 |m_{ll} - m_Z| > 15 \;{\rm GeV} \;{\rm and} \; \slashed{E}_T > 50\; {\rm GeV}.}
\label{eq:acuts}
\end{equation}
With these cuts at 8 TeV center-of-mass energy, the signal cross section for 150 GeV charged Higgs mass is
quite small ($\sim 0.13$ fb) whereas the background cross section is 64.12 fb. 
We note that the signal has additional sources of missing energy due to the presence of $\sigma$ particles 
which completely decay to neutrinos. Thus, selecting events with high missing transverse energy 
is expected to be helpful in distinguishing the signal from the background. However, when the mass of charged Higgs is close to 
$m_W$ the effect of large missing transverse energy cut is not very helpful.  As the mass of the 
charged Higgs increases, the fraction of events with higher transverse momentum ($p_{T}$) as well as 
higher missing transverse energy ($\slashed{E}_T$) is larger in the signal 
as compared to the background. Thus, the effect of harder missing energy cut becomes evident. This 
feature is illustrated in Fig.~\ref{fig:met}. 
The background, therefore, can be further reduced by raising 
the minimum missing energy cut. We find that the signal and background cross sections after applying 
100 GeV minimum $\slashed{E}_T$ cut become  0.04 fb and 2.11 fb respectively. This means, with the available 
luminosity of $\simeq 25\; {\rm fb}^{-1}$ at 8 TeV, for one signal event the number of 
background events is about 52. Therefore, it is very difficult to see the signal excess over such a 
large background at the 8 TeV LHC. The situation gets worse for $M_{H^\pm}$ = 200 GeV due to its smaller 
production cross section at 8 TeV. However, with larger center-of-mass energy (Fig.~\ref{fig:xsplot}) there is 
a significant increase in the pair production cross section.   
At $\sqrt{s}=14$ TeV the signal cross section is much larger and the data will be collected at much higher luminosity. 
Thus one expects to achieve greater signal significance at  $\sqrt{s}=14$ TeV run of the LHC.

\begin{figure}[ht!]
\includegraphics[width=3.15in]{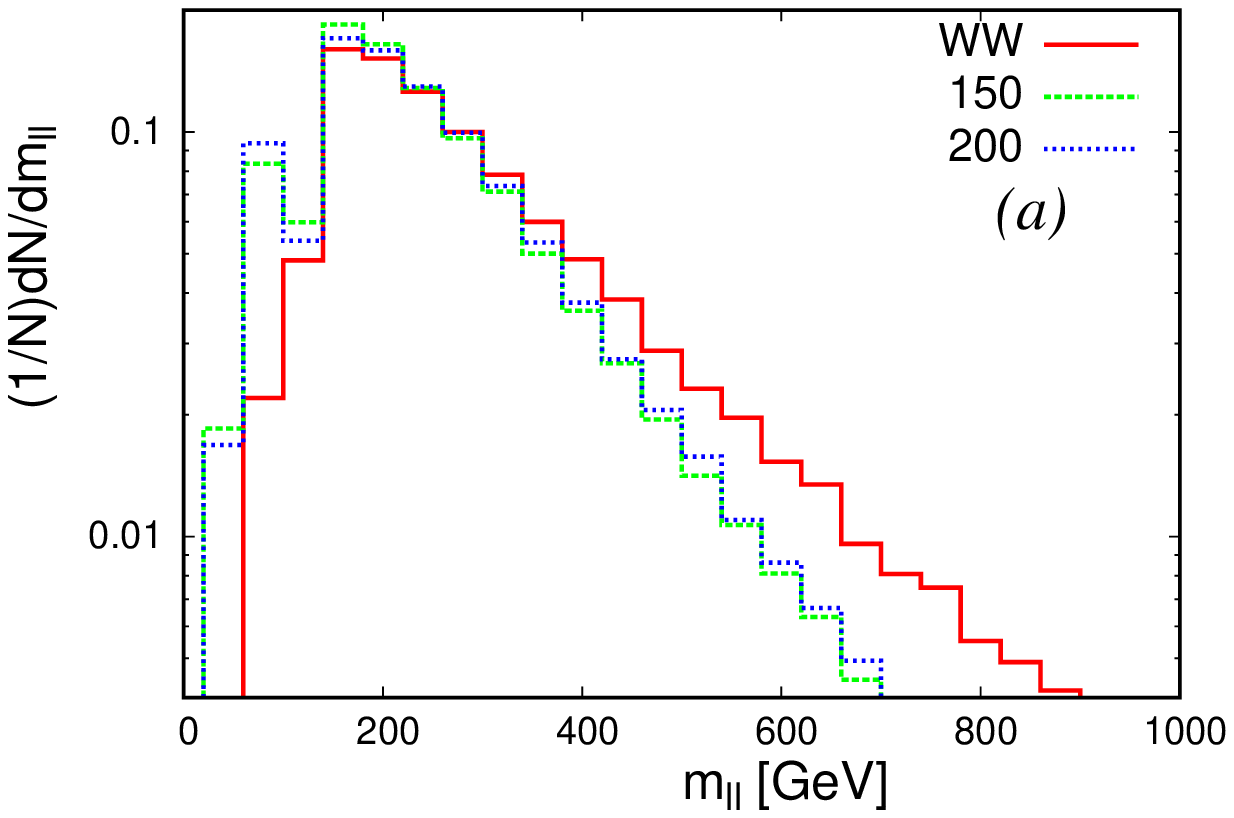}
\includegraphics[width=3.15in]{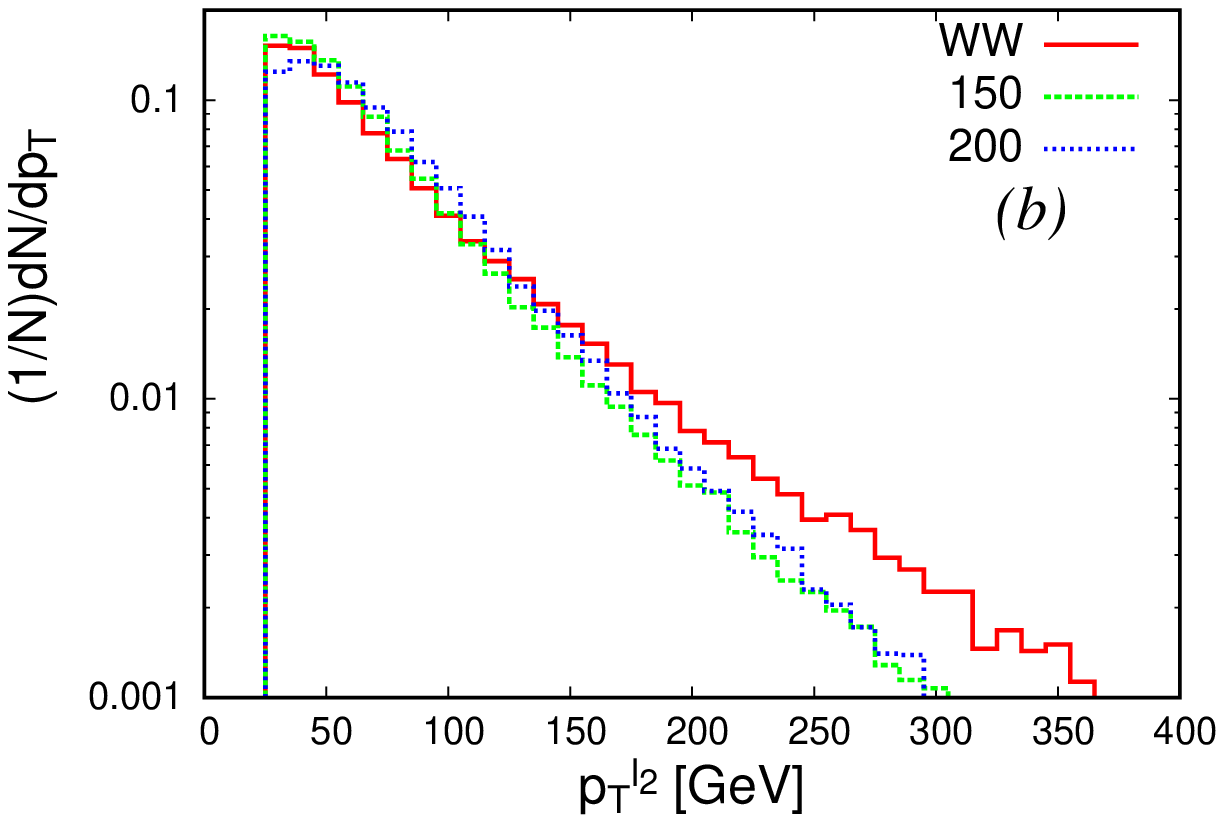}
\includegraphics[width=3.15in]{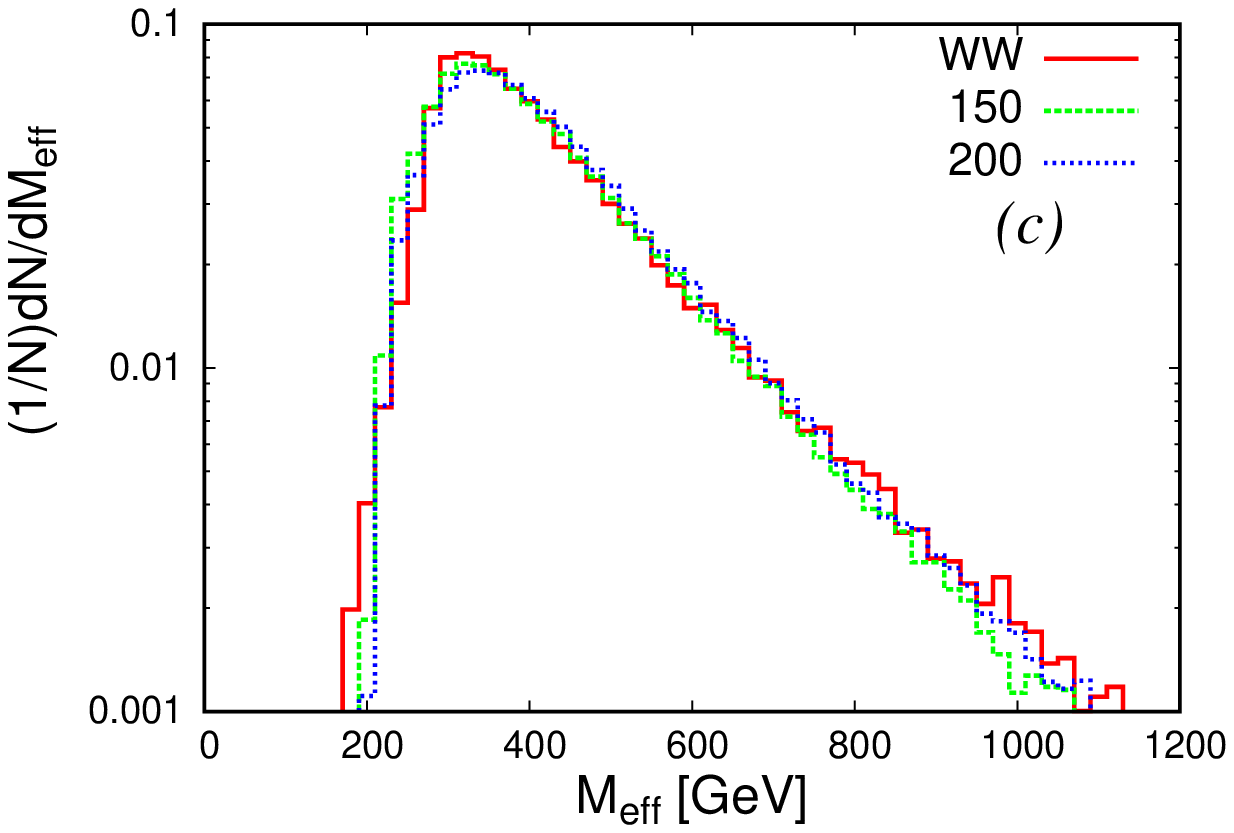}
\includegraphics[width=3.15in]{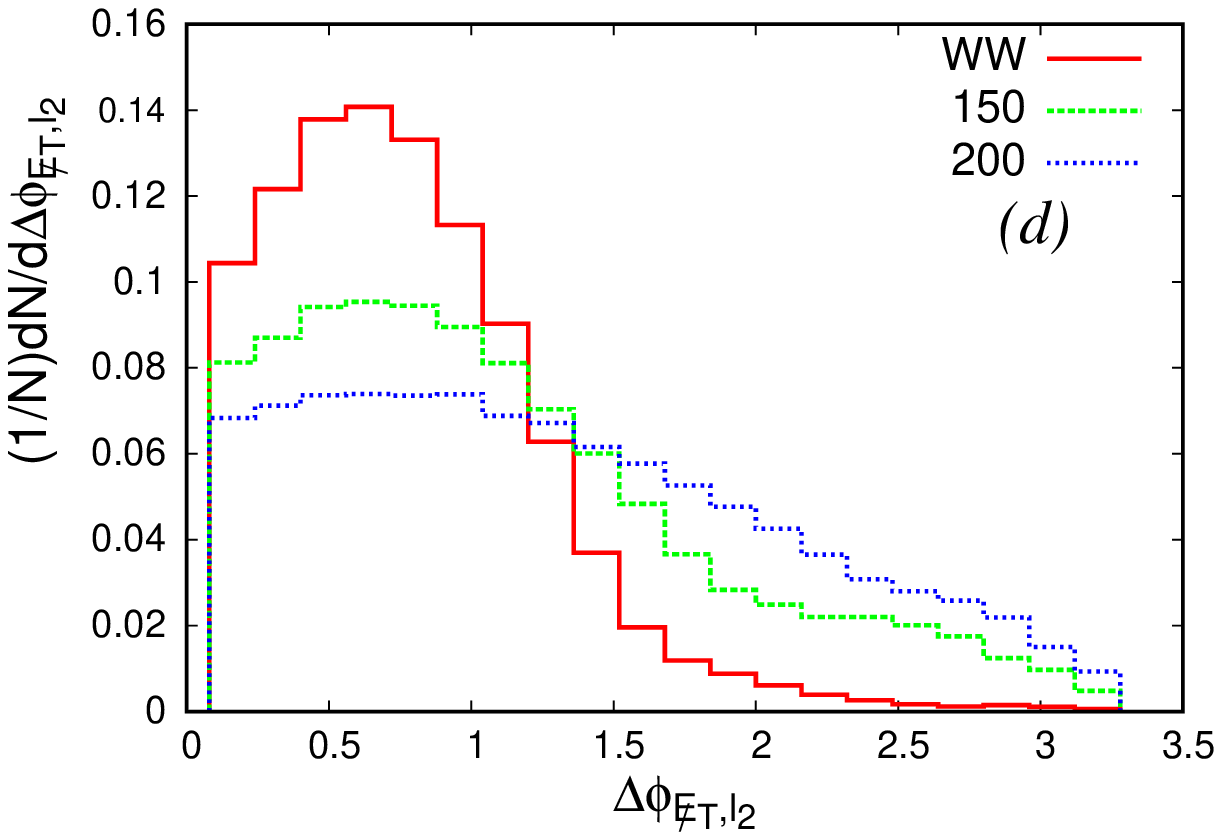}
\caption{Kinematic distributions for the ($2l+\slashed{E}_T$) signal with $M_{H^\pm}$=150, 200 GeV and background ($W^+W^-$). 
         The events satisfy the $\slashed{E}_T>110$ GeV cut and the acceptance cuts listed in Eq.~\ref{eq:acuts}.}
\label{fig:before_phi2}
\end{figure}
The event section criteria and basic acceptance cuts for the $\sqrt{s}=14$ TeV analysis are kept same, 
as shown in Eq.~\ref{eq:acuts}. 
The cross sections for the ($2l+\slashed{E}_T$) signal ($M_{H^\pm} = $ 150 GeV) and background are respectively, 0.25 fb and 96.21 fb after 
applying the acceptance cuts. Motivated by the observation in the $\sqrt{s}=8$ TeV analysis, we apply 
a minimum $\slashed{E}_T$ cut of about 110 GeV to enhance the signal significance to $\sim 2$. However, 
we note that raising the $\slashed{E}_T$ cut beyond 110 GeV does not improve the situation and we need 
to construct suitable kinematic variables which can help in reducing the background further. In Fig.~\ref{fig:before_phi2} 
we display kinematic distributions for the invariant mass ($m_{ll}$), the transverse momentum of the sub-leading lepton ($p_T^{l_2}$), 
the effective mass ($M_{eff}$)\footnote{$M_{eff}=\sum p_T^{visible}+\sum p_T^{missing}$ } and 
the angle between the directions of missing energy and the subleading lepton in the transverse plane ($\Delta \phi_{\slashed{E}_T,l_2}$). 
These distributions are plotted after applying the large $\slashed{E}_T$ cut. Quite clearly, it is the angle 
$\Delta \phi_{\slashed{E}_T,l_2}$ (Fig.~\ref{fig:before_phi2}d) which turns out to be the most effective kinematic variable in separating 
the signal from the background. We also note that this cut is more 
promising for the $M_{H^\pm}=200$ GeV. Based on this we have applied a minimum cut of 1.6 on the angle $\Delta \phi_{\slashed{E}_T,l_2}$. 
This improves the signal-to-background ratio significantly.

\begin{figure}[ht!]
\includegraphics[width=3.15in]{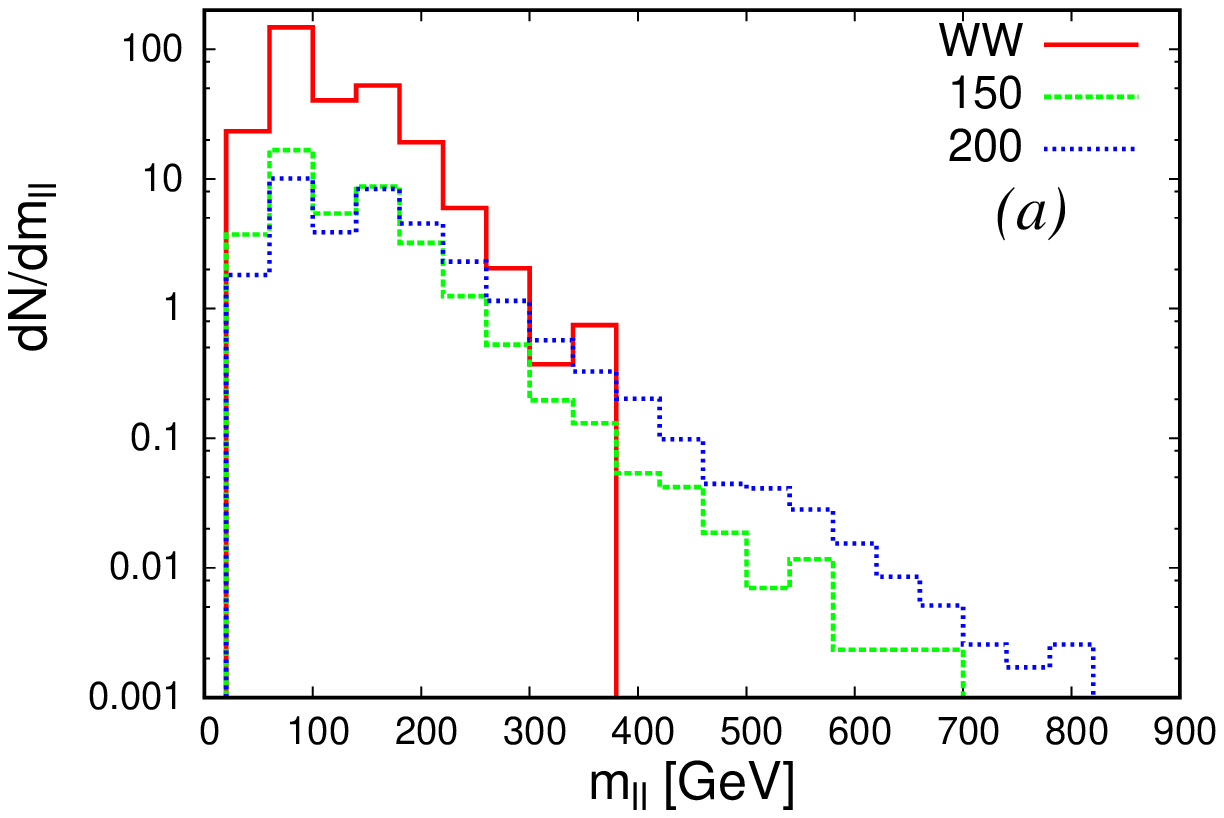}
\includegraphics[width=3.15in]{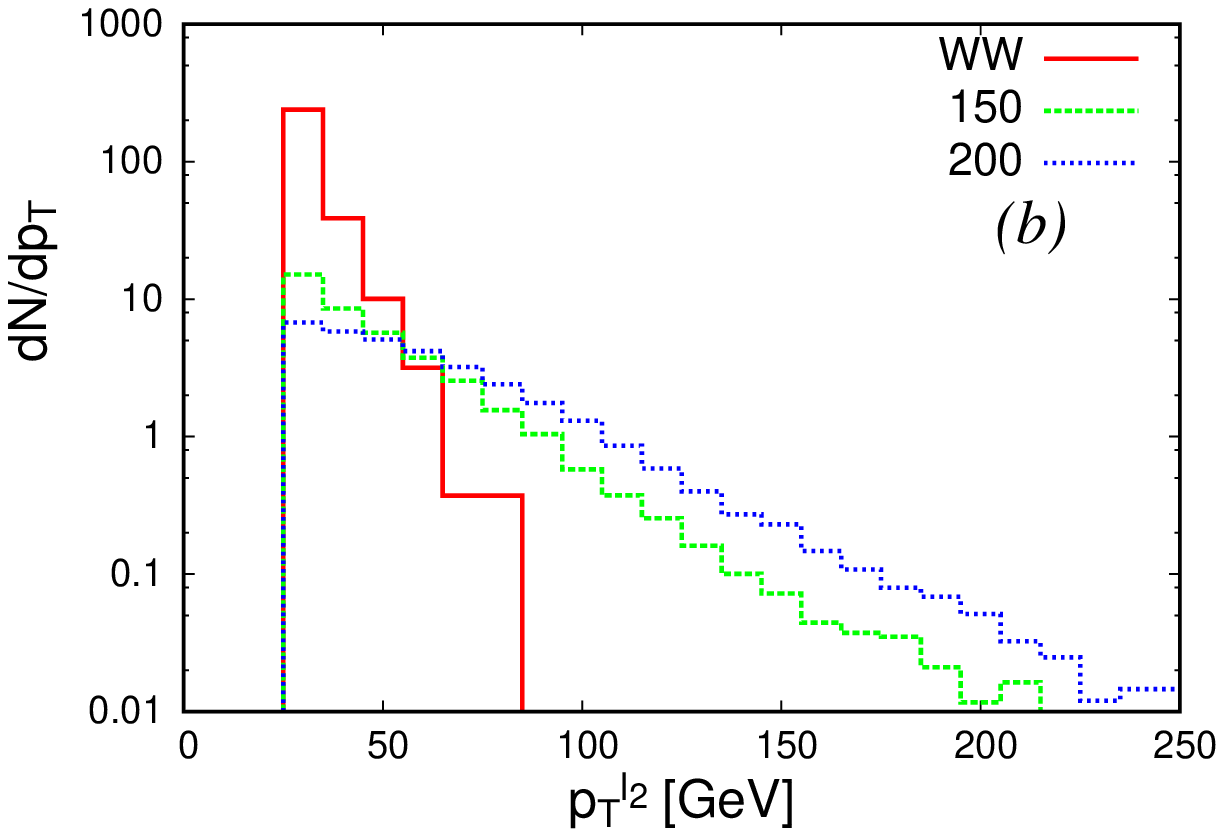}
\includegraphics[width=3.15in]{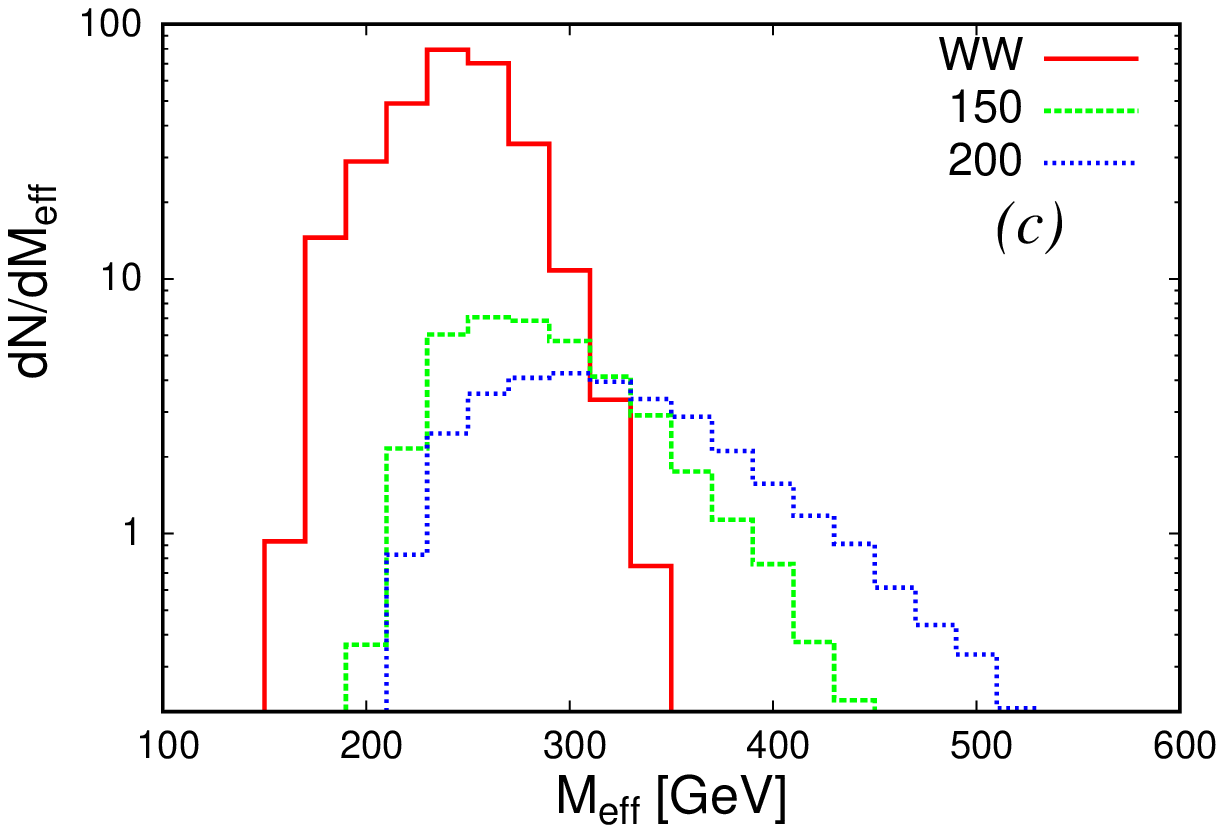}
\includegraphics[width=3.15in]{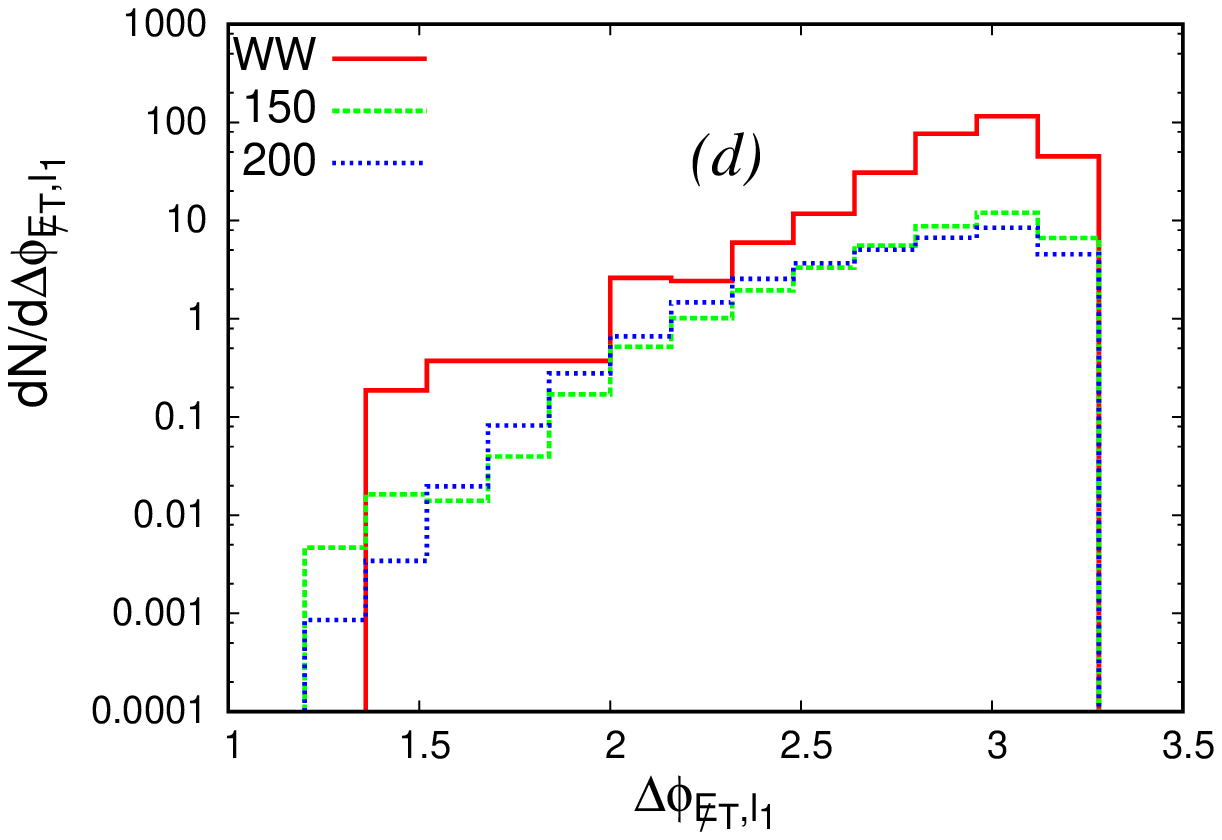}
\caption{Kinematic distributions for the ($2l+\slashed{E}_T$) signal  with $M_{H^\pm}$=150, 200 GeV and background ($W^+W^-$). 
         The events satisfy the $\Delta \phi_{\slashed{E}_T,l_2} > 1.6,~\slashed{E}_T>110$ GeV cut and the acceptance cuts 
listed in Eq.~\ref{eq:acuts}. }
\label{fig:after_phi2}
\end{figure}
 
The kinematic distributions shown in Fig.~\ref{fig:after_phi2} have been plotted after applying the 
$\Delta \phi_{\slashed{E}_T,l_2}$ cut. If we compare the distributions for $m_{ll}$, $p_T^{l_2}$  and
$M_{eff}$ in Fig.~\ref{fig:before_phi2} after the application of $\Delta \phi_{\slashed{E}_T,l_2}$ cut, as 
shown in  Fig.~\ref{fig:after_phi2}, we find that this cut affects the background events quite significantly.
In our case, the additional source of missing energy reduces the correlation
  between the leptons and the $\slashed{E}_T$, which is so crucially present in
  the $W^+W^-$ background. This causes events  of  background
  events with high $p_T$ to be removed once the above azimuthal angle cut is applied.
We can see from these distributions that the cut on the angle $\Delta \phi_{\slashed{E}_T,l_2}$ followed by a 
suitable cut on $p_{T}^{l_2}$ / $M_{eff}$ looks very promising in enhancing the signal significance.  The kinematic distribution displayed 
in Fig.~\ref{fig:after_phi2}d indicates that a minimum cut on $\Delta \phi_{\slashed{E}_T,l_1}$\footnote{Angle 
between the missing energy and the leading lepton in the transverse plane} may also help in improving the 
significance slightly. We have chosen a minimum cut of 55 GeV on $p_{T}^{l_2}$ and a minimum cut of 1.8 on 
the angle $\Delta \phi_{\slashed{E}_T,l_1}$ in our present analysis. With the help of these optimal values of cuts, 
a signal significance of about 4.6 can be achieved assuming an integrated luminosity $L = $ 3000 fb$^{-1}$, for the 
case of $M_{H^\pm}=150$ GeV. 

\begin{table}[ht!]
 \begin{center}
  \begin{tabular}{|c|ccc|cc|cc|}
   \hline
   {\bf Cuts applied} &\multicolumn{3}{c|}{\bf No of events}
   &\multicolumn{2}{c|}{\bf S/B} & \multicolumn{2}{c|}{{\bf Significance}  ($S_{\sigma}$)}   \\ \cline{2-4}
                      &\multicolumn{2}{c|}{\bf $H^{+}H^{-}$(S)} & {\bf $W^{+}W^{-}$}{\bf (B)} & & & & \\ 
                       & \multicolumn{1}{c}{$M_{H^\pm}$=150} &  \multicolumn{1}{c|}{(200) GeV}&  & $M_{H^\pm}$=150 & (200) GeV &$M_{H^\pm}$=150 & (200) GeV  \\
   \hline\hline
   Initial Signal                          & \multicolumn{1}{c}{2331.8}& \multicolumn{1}{c|}{(855.3)} & 931500.0 & 0.002 & (0.001) & 2.3 & (0.9) \\
   Isolation + 0j                          & \multicolumn{1}{c}{943.7} & \multicolumn{1}{c|}{(321.3)} & 431865.4 & 0.002 & (0.001) & 1.4 & (0.5) \\
   Acceptance cut                          & \multicolumn{1}{c}{739.5} & \multicolumn{1}{c|}{(263.9)} & 288626.8 & 0.002 & (0.001) & 1.4 & (0.5) \\
   $\slashed{E}_T >110$ GeV                & \multicolumn{1}{c}{201.6} & \multicolumn{1}{c|}{(107.9)} & 7423.1   & 0.028 & (0.014) & 2.3 & (1.2) \\
  $\Delta \phi_{\slashed{E}_T,l_2} > 1.6$  & \multicolumn{1}{c}{40.0}  & \multicolumn{1}{c|}{(33.5)}  & 292.1    & 0.130 & (0.115) & 2.3 & (1.9) \\
  $p_{T}^{l_2} > 55$  GeV                  & \multicolumn{1}{c}{8.6}   & \multicolumn{1}{c|}{(13.6)}  & 1.9      & 4.382 & (7.301) & 4.3 & (6.2) \\	  
 $\Delta \phi_{\slashed{E}_T,l_1} > 1.8$   & \multicolumn{1}{c}{8.5}   & \multicolumn{1}{c|}{(13.5)}  & 1.5      & 5.415 & (9.020) & 4.6 & (6.5) \\
  \hline
  \end{tabular}
 \end{center}
 \caption{Cut flow table at $14$ TeV center-of-mass energy and $3000\; {\rm fb}^{-1}$ integrated luminosity for $M_{H^\pm} = $150 GeV and $200$ GeV. 
          The significance ($S_\sigma$) is defined in Eq.~\ref{eq:significance}.}
\label{Table:150}
\end{table}

The effects of applying various cuts on the signal and background events have been summarized 
as a cut flow scheme in Table~\ref{Table:150} for the two benchmark values of the charged Higgs mass of 150 GeV and 200 GeV. 
We have selected only those cuts that increase the signal significance $(S_{\sigma})$ defined as 
\begin{equation}\label{eq:significance}
 S_{\sigma} = \sqrt{2(S+B){\rm ln}(1+S/B) - 2 S },
\end{equation}
where $S$ and $B$ are number of signal and background events respectively.
This significance estimator is useful for events with low statistics~\cite{Cowan:2010js}. 
When the background is large, the formula for $S_\sigma$ reduces to the more 
familiar $S/\sqrt{B}$ form used for estimating the signal significance.
Although the number of events that satisfy all the applied cuts are low, the significance is nevertheless promising. With the 3000 fb$^{-1}$ integrated luminosity, 
the same set of cuts lead to a signal significance of about 6.5 for $M_{H^\pm}=200$ GeV. 
We can further improve the significance by optimizing various cuts. 
For example, pushing the minimum $\slashed{E}_T$ cut on the higher side does help in achieving better significance.
As mentioned before, a suitable large cut on $M_{eff}$ instead of the cut on $p_{T}^{l_2}$ can also be used to suppress the background efficiently. 
However, in that case the minimum $\slashed{E}_T$ cut should be relaxed slightly to maintain a high signal significance.

\begin{figure}[t]
\includegraphics[width=3.00in,height=3.00in]{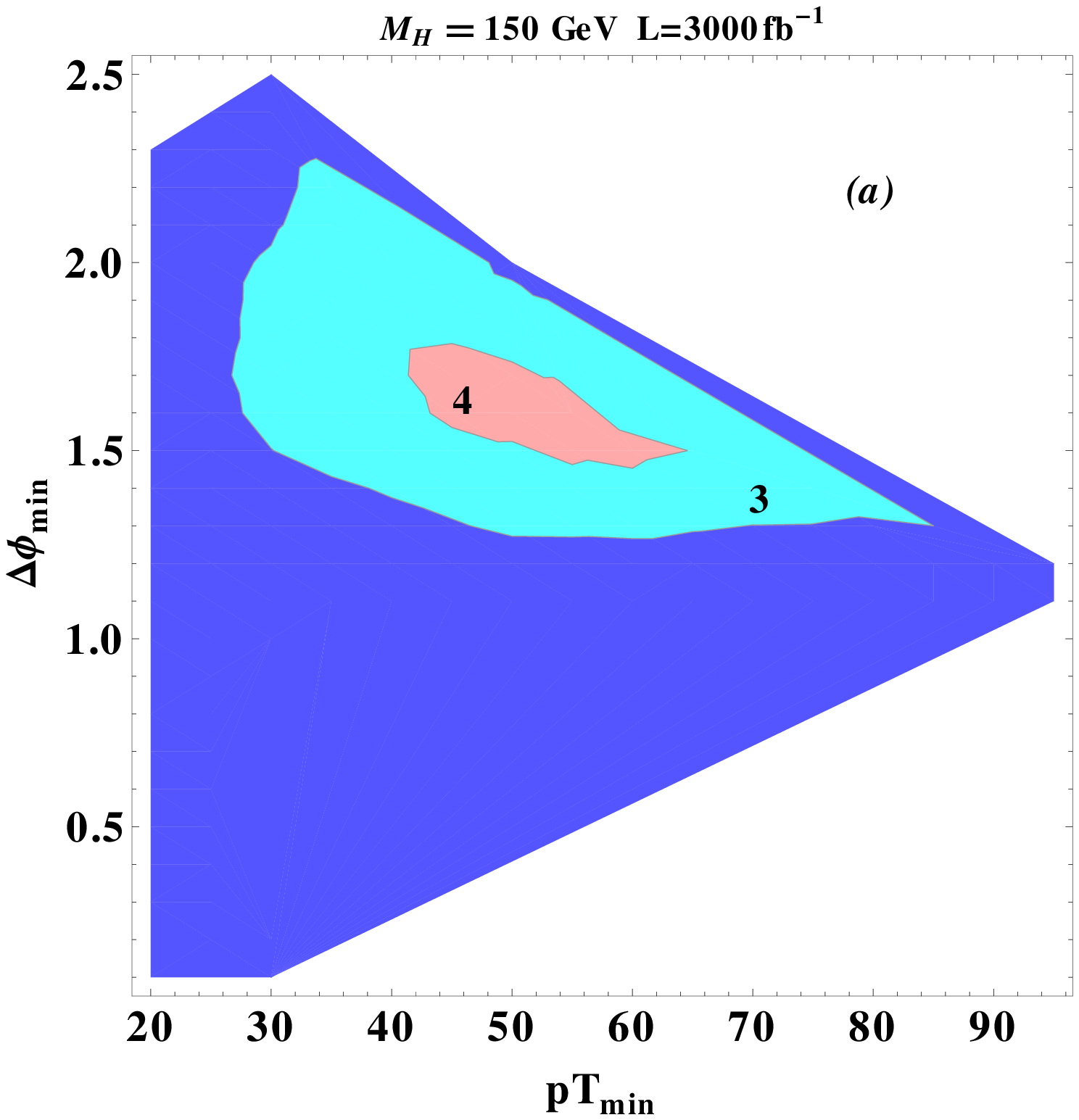}
\includegraphics[width=3.00in,height=3.00in]{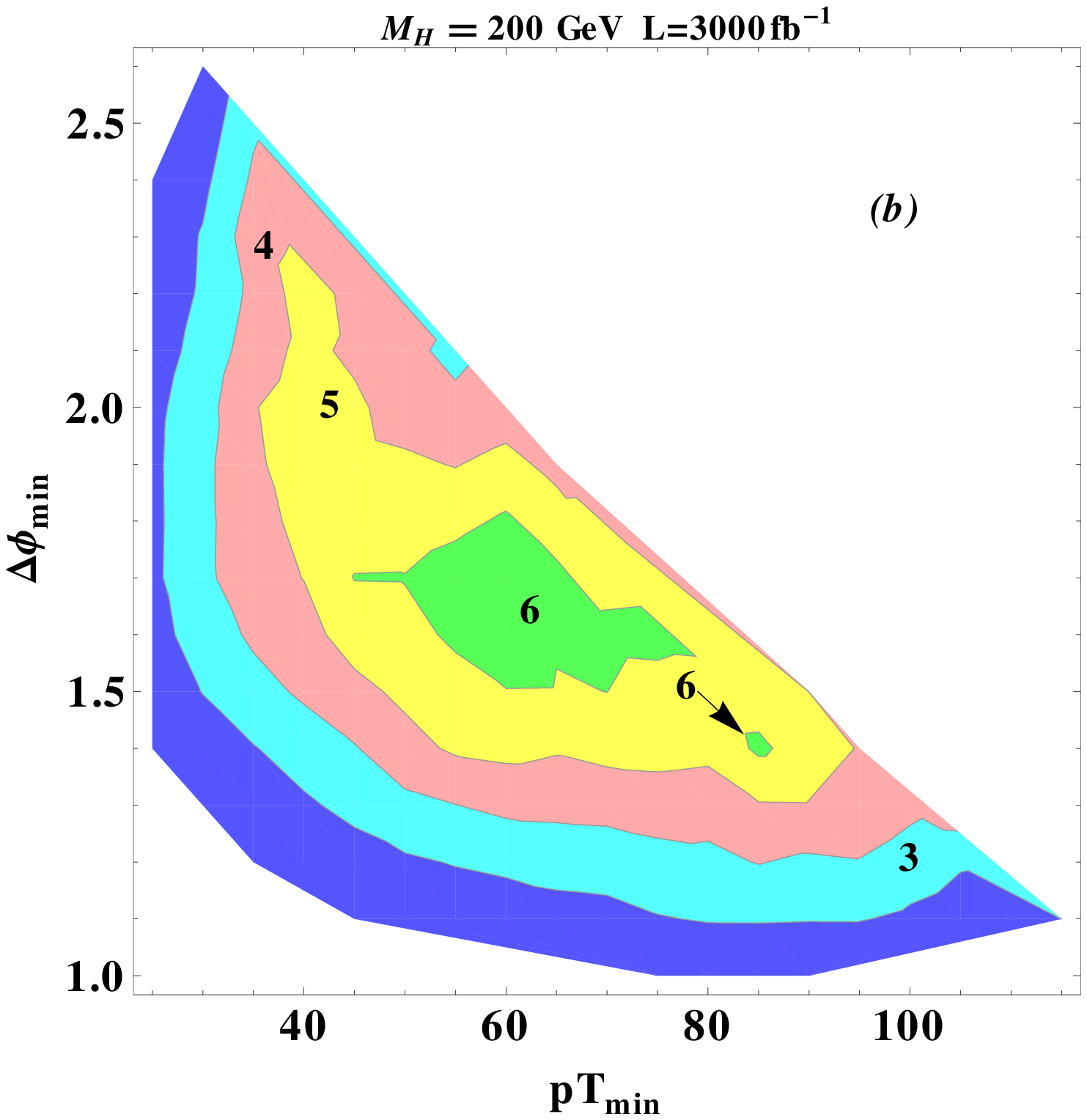}
\includegraphics[width=3.00in,height=3.00in]{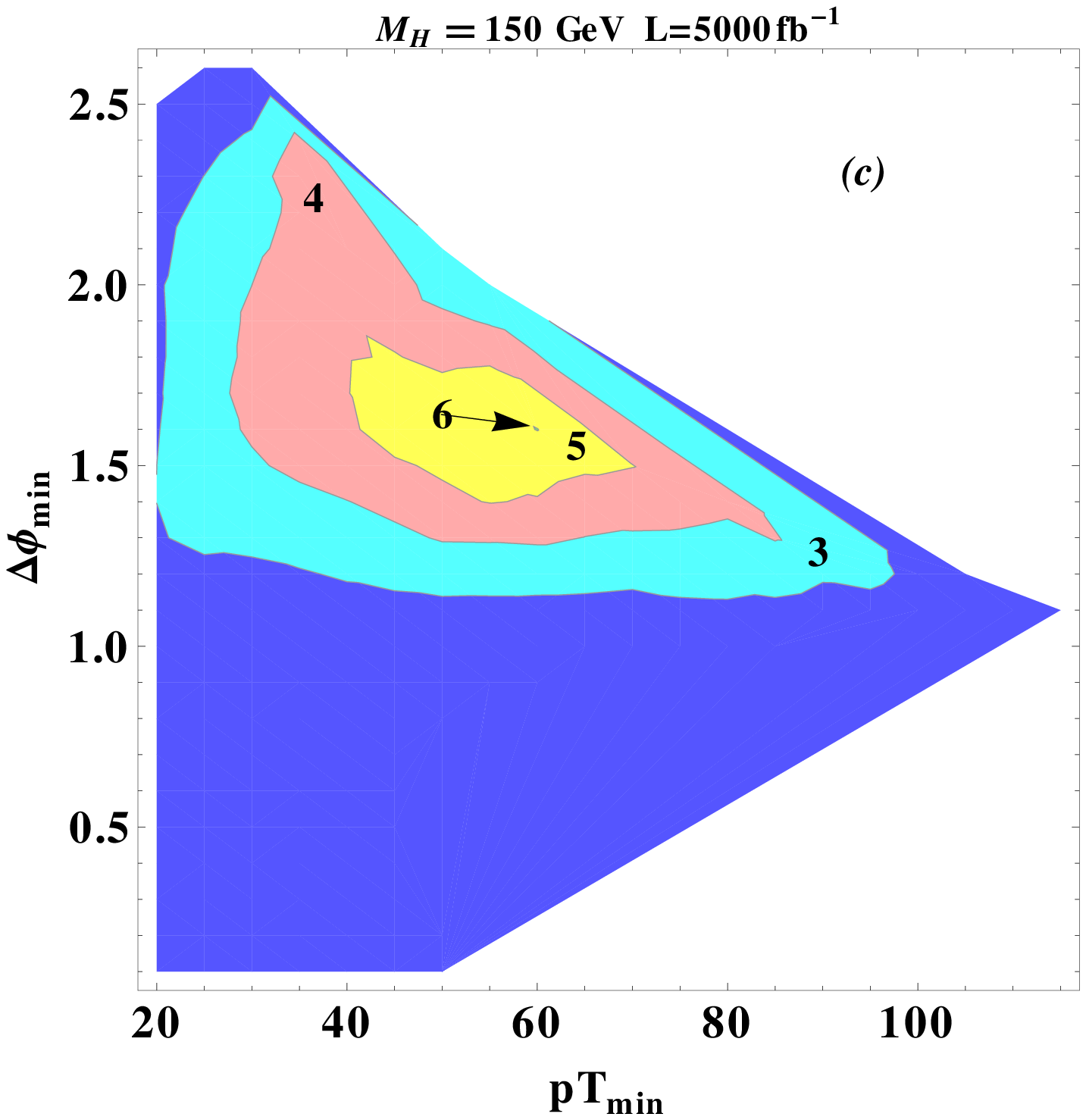}
\includegraphics[width=3.00in,height=3.00in]{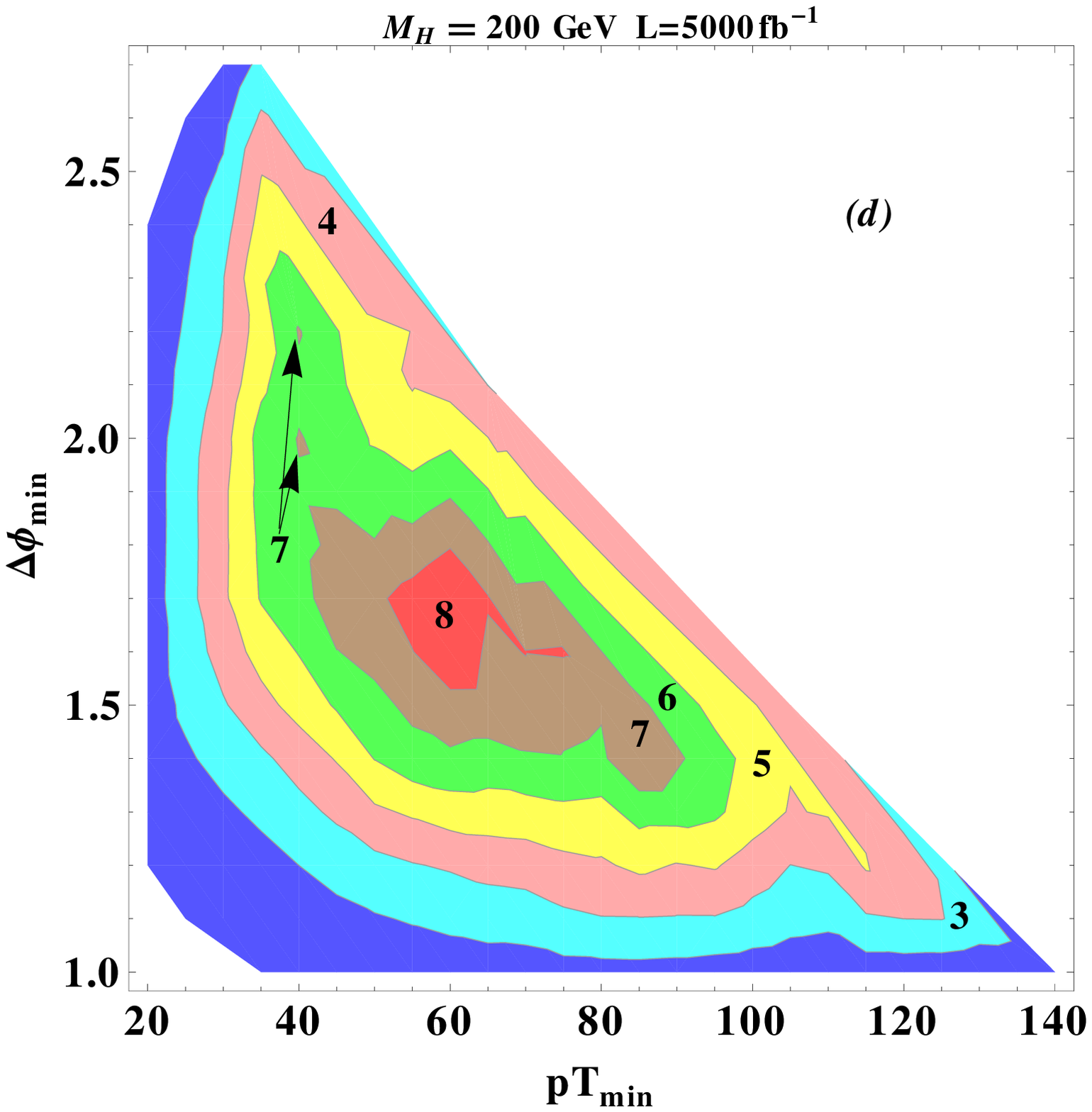}
\caption{Contour plots for the significance ($S_\sigma$) as a function of minimum cuts on $\Delta \phi_{\slashed{E}_T,l_2}$ (y-axis) and $p^{l_2}_{T}$ (x-axis) for (a) $M_{H^\pm}=150~{\rm GeV},~L=3000$ fb$^{-1}$, 
(b) $M_{H^\pm}=200~{\rm GeV},~L=3000$ fb$^{-1}$, (c) $M_{H^\pm}=150~{\rm GeV},~L=5000$ fb$^{-1}$, and, (d) $M_{H^\pm}=200~{\rm GeV},~L=5000$ fb$^{-1}$. The blue shaded regions in the above plots refer to 2$\sigma$ statistical significance.}
\label{fig:contours}
\end{figure}

As both $\Delta \phi_{\slashed{E}_T,l_2}$ and $p_{T}^{l_2}$ play an important role in enhancing 
the signal significance, it will be useful to study a possible correlation between the minimum cuts that 
can be applied on these kinematic variables. We have shown this correlation in Fig.~\ref{fig:contours} using the 
contour plots for $M_{H^\pm}$=150, 200 GeV. The plots are shown for two values of the integrated luminosity,  {\it viz} $L=3000$ fb$^{-1}$ and $L=5000$ fb$^{-1}$. The contour plots have been obtained after 
applying the minimum $\slashed{E}_T>110$ GeV cut along with the acceptance cuts. In these plots we show 
the minimum cuts on $p_{T}^{l_2}$ and $\Delta \phi_{\slashed{E}_T,l_2}$ required to achieve a signal significance $S_{\sigma} \ge$ 2, 
keeping the total number of events $S+B \ge$ 5. The signal significance is clearly seen to increase substantially 
with more optimal choices of cuts for the two correlated kinematic variables. The variation in the signal 
significance is represented with different color codes. With the help of these plots it is easier to find the 
optimal values of the cuts on $\Delta \phi_{\slashed{E}_T,l_2}$ and $p_{T}^{l_2}$ which can maximize 
the significance. We should also point out that our analysis for the two benchmark values of the charged Higgs 
mass achieved a signal significance in accordance with the contour plots shown in Fig.~\ref{fig:contours}a and  \ref{fig:contours}b. 
 However, the efficiency of these correlated cuts for the case of $M_{H^\pm}=200$ GeV which
is clearly visible in Fig.~\ref{fig:contours}b, suggests that a significance, as high as 5$\sigma$, should
be well within the reach with more optimal choices of the cuts when compared to those 
listed in Table~\ref{Table:150}. As expected, with higher integrated luminosity, much better signal 
significance can be achieved (see Figs.~\ref{fig:contours}c and \ref{fig:contours}d). It is worth pointing out
that in Fig.~\ref{fig:contours}, the constraint $S+B \ge$ 5 plays a major role in modifying the shape of  
common significance contours. As the discrete cut-off for events will not be uniform for both $L =$ 3000 fb$^{-1}$ 
and $L = $ 5000 fb$^{-1}$, it gives an impression of a non trivial scaling at different luminosities.  
 
 \begin{figure}[ht!]
\includegraphics[width=4.2in]{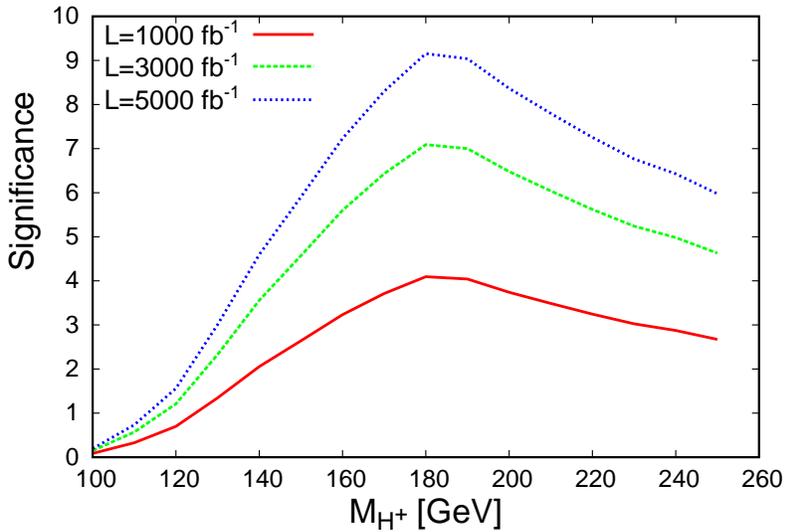}
\caption{Illustrating the signal significance for different charged Higgs masses and integrated luminosities. The kinematic cuts are the same as given in Table~\ref{Table:150}.}
  \label{fig:discovery}
\end{figure}
 
In Fig.~\ref{fig:discovery}, 
we estimate the signal significance for various charged Higgs masses assuming 1000, 3000 and 5000 fb$^{-1}$
integrated luminosities at the LHC with $\sqrt{s} = 14$ TeV. We have applied the same set of cuts as listed in Table~\ref{Table:150}.
Since we have applied a large missing energy cut, the significance increases for higher charged Higgs 
masses. However, beyond a certain value of charged Higgs mass, the significance goes down. This is mainly due 
to the small pair production cross section of the charged Higgs. We note that for $M_{H^\pm}=180$ GeV 
the significance is seen to become maximum in Fig.~\ref{fig:discovery}. This is just the artifact of the choice of
$\slashed{E}_T$ cut, which is more effective at that charged Higgs mass. Note that beyond the charged Higgs mass of 
about 180 GeV, the $H^\pm \to W^\pm \rho$ decay channel is also open for $M_\rho =$ 100 GeV. This reduces the charged Higgs 
branching ratio in the $W\sigma$ decay mode and hence reduces the signal cross section further for $M_{H^\pm} > $ 180 GeV. 
The upper bound on $M_{\rho}$ which is related to an upper bound on the coupling $\lambda_5$ can be 
about 470 GeV~\cite{Gabriel:2006ns}. It means, by considering heavier $\rho$ mass we can ensure 
a 100\% branching ratio of the charged Higgs decay to $W^\pm \sigma$ for the heavier charged Higgs masses.  
It is needless to say that optimization of kinematic cuts is required to estimate the actual signal significance 
for different charged Higgs masses. For example, despite larger signal cross section for $M_{H^\pm} = 120$ GeV, 
the significance is maximized for a minimum $\slashed{E}_T$ cut of 90 GeV. On the other hand, for $M_{H^\pm} = 220$ GeV 
we have smaller signal cross section but a minimum cut of 130 GeV on 
$\slashed{E}_T$ is more helpful in achieving a larger signal significance. We note that ($\Delta\phi_{\slashed{E}_T,l_2},p_{T}^{l_2}$) 
combination which we have used to discriminate the 
signal from the background for $M_{H^\pm} = 150,~200$ GeV is not very useful in increasing the significance 
for lower $M_{H^\pm}$ values. This is related to the fact that the efficiency of these cuts is closely 
related to the high $\slashed{E}_T$ cut which in turn is related to the mass difference between the charged Higgs and the $W$ boson. 
When the mass difference between the charged Higgs and the 
$W$ boson is not large enough, the application of high $\slashed{E}_T$ cut also kills the signal along with 
the background. It is also worth pointing out that for larger charged Higgs masses a suitable 
$\Delta \phi_{\slashed{E}_T,l_2}$ cut with large enough $\slashed{E}_T$ cut is sufficient. In other words, if 
it is possible to apply a very high $\slashed{E}_T$ cut, the additional $p_{T}^{l_2}$ cut becomes less 
relevant.
\section{Summary and Conclusion}
 \label{sec:summary}
We have studied the signatures of charged Higgs boson at the LHC in a two Higgs doublet
model with right-handed neutrinos.  The model, aiming to derive
neutrino masses via preferential Yukawa couplings $\cal O$(1) with an additional 
Higgs doublet, also implies large Yukawa coupling of the charged Higgs with the light 
leptons and neutrinos. However, if cosmological constraints are taken into account, the 
leptonic decay mode of the charged Higgs is highly suppressed and $H^\pm \to
W^\pm \sigma$ is the dominant decay mode. In this study, the charged
Higgs pair production via the Drell-Yan process and its further decay leading to 
opposite sign di-leptons+missing energy in the final state is considered as the signal. 
The major SM background to the signal comes from the process, $pp \to W^+ W^-$. 
We have done a complete signal-background cut based analysis for both the 8 TeV and 
14 TeV center-of-mass energies at LHC. The charged Higgs masses of 150 GeV and 
200 GeV serve as the benchmark points for our study.  Since the signal has additional 
sources of missing energy, we find that a large $\slashed{E}_T$ cut helps in suppressing 
the SM background. Also, a combination of minimum cuts on the angle 
$\Delta\phi_{\slashed{E}_T,l_2}$ and $p_{T}^{l_2}$ plays an important role in 
enhancing the signal significance. Due to the lack of sufficient data and the low signal 
cross section as compared to the background,  the observability of the signal is not 
possible at the LHC with $\sqrt{s}= 8$ TeV and we therefore carry out our analysis for 
the 14 TeV run of the LHC. We find that even at the 14 TeV run of LHC, a charged Higgs 
with the characteristics of a ``fermiophobic" field will prove elusive without 
a very optimized kinematic selection of events even with high integrated luminosity. We 
show this by identifying the kinematic variables sensitive to specific selection 
cuts which with a large (3000 fb$^{-1}$) integrated luminosity yields a signal significance 
of 4.6$\sigma$ for $M_{H^\pm}=$ 150 GeV. For the case of $M_{H^\pm}=$ 200 GeV, a 
signal significance of 5$\sigma$ can be easily achieved with better statistics. We highlight 
the significance of the optimized cuts in the analysis through a correlation plot for 
event selection that enhances the signal significance in a very robust way. Our analysis 
indicates that the observation of this otherwise elusive fermiophobic charged Higgs 
boson is quite promising at the high energy and high luminosity run of the 
LHC, provided a proper event selection criterion is applied.
\begin{acknowledgments}
We thank Arindam Chatterjee and Sandhya Choubey for fruitful discussions.
The work of  UM, BM, SKR and AS was partially supported by funding available from the Department of 
Atomic Energy, Government of India, for the Regional Centre for Accelerator-based Particle Physics, 
Harish-Chandra Research Institute. The work of SN was  was supported in part by the U.S. Department 
of Energy Grant Number DE-SC0010108.
\end{acknowledgments}
%

\end{document}